# STAN: Spatio-Temporal Attention Network for Pandemic Prediction Using Real-World Evidence

**Junyi Gao[1,4], Rakshith Sharma[2], Cheng Qian[1], Lucas M. Glass, [1,3] Jeffrey Spaeder, M.D. [1], Justin Romberg[2], Jimeng Sun[4], and Cao Xiao[1]**

[1]IQVIA, [2]Georgia Institute of Technology [3]Temple University and [4]University of Illinois at Urbana-Champaign

Correspondence to Cao Xiao, PhD, IQVIA, 201 Broadway Floor 5, Cambridge MA 02139, USA; Email: cao.xiao@iqvia.com and Jimeng Sun, PhD, 201 North Goodwin Avenue Urbana, IL 61801; Email: jimeng@illinois.edu

## ABSTRACT

**Objective**: We aim at developing a hybrid model for earlier and more accurate predictions for the number of infected cases in pandemics by (1) using patients' claims data from different counties and states that capture local disease status and medical resource utilization; (2) utilizing demographic similarity and geographical proximity between locations; and (3) integrating pandemic transmission dynamics into a deep learning model.

**Materials and Methods**: We proposed a Spatio-Temporal Attention Network (STAN) for pandemic prediction. It uses a graph attention network to capture spatio-temporal trends of disease dynamics and to predict the number of cases for a fixed number of days into the future. We also designed a dynamics-based loss term for enhancing long-term predictions. STAN was tested using both real-world patient claims data and COVID-19 statistics over time across the US counties.

**Results**: STAN outperforms traditional epidemiological models such as Susceptible-Infectious-Recovered (SIR), Susceptible-Exposed-Infectious-Recovered (SEIR), and deep learning models on both long-term and short-term predictions, achieving up to 87% reduction in mean squared error compared to the best baseline prediction model.

**Conclusions**: By combining information from real-world claims data and disease case counts data, STAN can better predict disease status and medical resource utilization.

# OBJECTIVE

Pandemic diseases such as the novel coronavirus disease (COVID-19) has been spreading rapidly across the world and poses a severe threat to global public health. Up to July 2020, COVID-19 has affected 14.1 million people and caused more than 597K deaths over the world[1], and caused significant disruption to people's daily life as well as substantial economic losses. Therefore, it is critical to predict the pandemic outbreak early and accurately to help design appropriate policies and reduce losses.

Many epidemiological models (e.g., susceptible-infected-removed (SIR), susceptible-exposed-infected-removed (SEIR)), and deep learning models (e.g., Long Short Term Memory networks - LSTM) have been applied to predict the COVID-19 pandemic [1-4]. However, they face three major challenges: (1) They usually build a separate model for each location (e.g., one model per county) without incorporating geographic proximity and interactions with nearby regions. Or the forecasts only depend on some observed patterns from other locations[2, 3], while inter-regional interactions are not directly modeled. In fact, a location often shows similar disease patterns with its nearby locations or demographically similar locations due to population movements or demographic similarity [5]. (2) Existing models are mainly built on COVID-19 case report data. These data are known to have severe under-reporting or other data quality issues. (3) Epidemiological models such as SIR and SEIR are deterministic models. They use a set of differential equations to fit the entire curve of disease counts. These models are determined by only a few parameters, making them unable to capture complex short-term patterns such as superinfection or time-varying infectivity[4]. Conversely, deep learning-based models can only predict known data patterns and lead to accurate predictions only within a short time period. Therefore, while there are techniques that allow for either short-term or long-term predictive models of disease outbreaks, existing models do not provide accurate models over both time horizons.

In this work, we propose a Spatio-Temporal Attention Network (STAN) for pandemic prediction using real-world evidence such as claims data and COVID-19 case surveillance data. We map locations (e.g., a county or a state) to nodes on a graph and construct the edges based on geographical proximity and demographic similarity between locations. Each node is associated with a set of static and dynamic features extracted from multiple real-world evidence in medical claims data that capture disease prevalence at different locations and medical resource utilization conditions. We utilize a graph attention network (GAT) to incorporate interactions of similar locations. Then we predict the number of infected patients for a fixed period into the future while concurrently imposing physical constraints on predictions according to transmission dynamics of epidemiological models. We apply STAN to predict both state-level and county-level future number of infected cases, achieving up to 87% reduction in mean squared error compared to the best baseline model. The study has been determined by the UIUC IRB as non-human subject research.

# BACKGROUND AND SIGNIFICANCE

## Epidemic/Pandemic Prediction Models

Traditional epidemic prediction models use compartmental based models that estimate disease transmission dynamics at the population level, such as SIR, SEIR models, and their variants[2, 3]. Some works also utilize time series learning approaches for pandemic prediction, for example, applying curve-fitting[3] or autoregression[5]. Besides these traditional statistical models, deep learning models were developed to cast epidemic or pandemic modeling as time series prediction problems. Many works [6-8] combines deep neural networks (DNN) with causal models for influenza-like illness (ILI) incidence forecasting. Deng *et al.*[9] proposed a graph message passing framework to combine learned feature embeddings and an attention matrix to model disease propagation over time. Yang *et al.*[3] used previous pandemic data to pretrain the LSTM, and then apply it to predict COVID-19 progression in China. Kapoor *et al.*[10] utilize a simple graph neural network for COVID-19 prediction. However, these models only predict the next day instead of long-term progression. It is still challenging to make deep-based models achieve good long-term prediction performance. Moreover, DNN-based methods have a significant issue: they can only predict known trends from the input data without understanding the long-term progression trend. For example, at the early stage of the pandemic if all case counts are increasing, it is unlikely for these models to predict a declining trend in future. Hence the long-term prediction is often difficult for DNN models.

## Incorporate Disease Transmission Dynamics in Graph Neural Network

Recently, several studies have attempted to incorporate knowledge about physical systems into deep learning. For example, Wu *et al.* and Beucler *et al.*[11, 12] introduced statistical and physical constraints in the loss function to regularize the model's predictions. However, their studies only focused on spatial modeling without temporal dynamics. Seo *et al.*[13, 14] integrates physical laws into graph neural networks. However, they focused on using physical laws to optimize node-edge transitions instead of concentrating on prediction results. In particular, those models only predict graph signals for the next time point instead of long-term outcomes. In our work, we also incorporate physics laws, i.e., disease transmission dynamics, to regularize model predictions to overcome the limitations of the prior models. These regularizations will be applied over a time range to ensure we can predict long-term pandemic progression. Since our proposed method is applied to extracted temporal and spatial embeddings of locations as an extra loss term, it does not introduce extra hyper-parameters; hence it is easier to train.

## MATERIALS AND METHODS

### Problem Formulation

In this paper, we develop the model STAN to predict the number of COVID-19 cases for a fixed number of days into the future, at a county or state level across the USA. STAN takes the following input data: county-level historical daily numbers of positive cases, county-level population-related statistics, and the frequencies of relevant medical codes extracted from medical claims data. Our goal is to better predict the number of cases by utilizing the rich information captured by these different data sources.

Throughout the paper, we use $N$ to denote the number of spatial locations (counties), $X$ to represent the feature matrix of size $N \times (F_S + T \times F_D)$ where $F_S$ is the number of *static* features per county (or state) and $F_D$ is the number of *dynamic* features for each county (or state). $T$ denotes the total number of time steps (i.e., days) for each location. Finally, we are interested in predicting $I(t)$, the number of infected patients at the $t^{th}$ time step for all the locations.

As depicted in **Figure 1**, STAN is enabled by the following components: 1) a graph neural network that captures the geographic trends in disease transmission; 2) an RNN that captures the temporal disease patterns at each location; 3) Both short-term prediction loss and long-term transmission dynamics constraint loss to regularize learned hidden representations of node embeddings. We describe each of these aspects below.

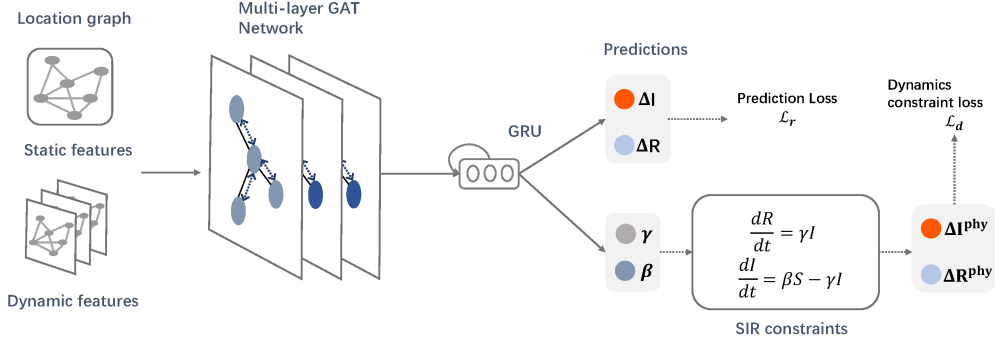

**Figure 1.** The STAN model: We construct the location graph using location-wise dynamic and static features as nodes and geographic proximity as edges. The graph is fed into graph attention network to extract spatio-temporal features and learn the graph embedding for the target location. Then the graph embedding is fed into the GRU to extract temporal relationships. The hidden states of GRU will be used to predict future number of infected and recovered cases. We use an additional transmission dynamics loss based on pandemic transmission dynamics to optimize the model.

## (I). Graph Construction

The input data includes dynamic data and static data. Dynamic data is a 3D tensor that includes location (e.g., states, counties, etc.), timestamp (e.g., days/weeks), and the dynamic features at each location (e.g., the number of active COVID-19 cases and the numbers of other related ICD codes). Static data is a 2D matrix that includes location and the static features for each location. We also form an **attributed graph** to capture the spatio-temporal epidemic/pandemic dynamics. In particular, we model the geographic proximity and demographic similarity between the different locations as edges in a location graph.

**Graph nodes:** We construct an attributed graph $G(\mathcal{V}, \mathcal{E})$ to represent the input data. A location is modeled as a graph node and is associated with a feature matrix that contains both static and dynamic features across all the time stamps for that location. In total, we have 193 nodes for the county level graph (one for each county with more than 1,000 infected cases) or 45 nodes for the state level graph (one for each state).

**Graph edges:** The edges are constructed based on geographical proximity and the population sizes of the nodes (i.e., locations). In particular, we designate the weight of an edge between nodes $i$ and $j$ as $w_{ij} \propto p_i^\alpha p_j^\beta \exp\left(-\frac{d_{ij}}{r}\right)$ where $p_i$ and $p_j$ are the population size of the nodes, $d_{ij}$ is the geographical distance between them, $\alpha, \beta$ and $r$ are hyperparameters. The above model is based on the idea that disease transmission patterns highly depend on crowd mobility. If there is a high mobility rate between a pair of nodes, we expect that the nodes have the similar disease spread parameters. Hence, our process includes an edge with a large weight between such a pair of nodes. Note that the distance parameter $d_{ij}$ can incorporate any general notion of distance, including inverse of the volume of air and car travels.

**Node features (static):** Each node $n_i$ has an associated static feature vector of size 4, consisting of the static features including latitude, longitude, population size, and population density.

**Node features (dynamic)** – Each node $n_i$ also has a set of dynamic features in the form of a matrix. The dynamic features include the number of active cases, total cases, the current number of hospitalizations and ICU stays due to COVID-19, which are calculated by aggregating the related procedure codes. We also include the number of each of the 48 COVID-19 related diagnosis codes extracted from claims data according to the Centers for Disease Control and Prevention guideline (https://www.cdc.gov/nchs/data/icd/COVID-19-guidelines-final.pdf). We outline the specific diagnosis codes used in the description of the dataset.

## (II). Modeling Spatio-temporal Patterns using Graph Attention Networks

Obtaining the complex spatial dependencies is a crucial problem to pandemic prediction. By utilizing spatial similarity, our model can make more accurate predictions of a location by considering similar locations' disease transmission status. Here we employ the Graph Attention Networks (GAT) model[15] to extract spatio-temporal similarity features. The basic idea of GAT is updating the embedding of each node by aggregating its neighboring nodes. In our setting, each location will receive information from its adjacent locations based on mobility to model spatio-temporal disease transmission patterns. This consideration is based on the real-world scenario that adjacent locations may have different impacts on the infectious status of the focused location. For example, if one city has a large population size and increasing infected cases, this city may have a more considerable impact on its adjacent counties.

We use a two-layer GAT to extract spatio-temporal features from the attributed graph. We use the latest values from historical data within a sliding window to construct the graph. Mathematically, at time step $t$, the input features to node $i$ are $X_t^i$, where $X_t^i \in \mathbb{R}^{L_I(F_D+F_S)}$ and $L_I$ denotes the length of the input window. Intuitively at the $t$-th timestep, we concatenate historical features (i.e., $L_I$ days of features) as input. The longer the $L_I$ is, the more historical information and patterns the model will use. Then we apply the graph attention mechanism and calculate the node representation $z_t^i \in \mathbb{R}^{F_z}$ for each node, where $F_z$ denotes the output dimension of the GAT layer.

Concretely, to calculate the $z_t^i$, we use the multi-head mechanism to calculate $K$ independent attention scores following the self-attention strategy[16]. A multi-head attention mechanism can help the model get more accurate predictions by generating different attention weights. Intuitively different attention heads may focus on different features in the graph to more comprehensively model the locations. The attention weight of the $k$-th head between two nodes $i$ and $j$ as:

$$e_{ij}^k = \sigma(\mathbf{W}_a^k(\mathbf{W}_z^k X_t^i | \mathbf{W}_z^k X_t^j))$$

where $\mathbf{W}_z^k \in \mathbb{R}^{F_z \times |X_t^i|}$ denotes the linear transformation weight matrix for the $k$-th head, which will transform the input to the output dimension. $\mathbf{W}_a^k \in \mathbb{R}^{1 \times 2F_z}$ represents the attention computation matrix for the $k$-th head, $(\cdot | \cdot)$ denotes the vector concatenation. $\sigma$ is the non-linear

activation function, and here we use the leaky rectified linear function (LeakyReLU):

$$\sigma(x) = \begin{cases} x, if\ x \geq 0 \\ 0.01x, otherwise \end{cases}$$

which is the same as the original GAT model[15].

Next we use the softmax function to calculate the attention score:

$$a_{ij}^k = \text{softmax}(e_{ij}^k) = \frac{\exp(e_{ij}^k)}{\sum_{n=1}^{N} \exp(e_{in}^k)}$$

Each edge of node $i$ will receive an attention score, which assesses how much information should be aggregated from neighboring node $j$. Finally, we sum up all embedding vectors from multiple heads to obtain the final representation $\mathbf{z}_t^i$ for node $i$ as:

$$\mathbf{z}_t^i = \frac{1}{K} \sum_{k=1}^{K} \sum_{j=1}^{N} a_{ij}^k \mathbf{W}_z^k \mathbf{X}_t^i$$

where $N$ denotes the number of locations.

## (III). Modeling Temporal Features using Recurrent Neural Networks

Pandemic prediction is not only spatial-related, but also a temporal-related task. The graph information will change along with time. We want to model the spatio-temporal patterns to better predict future trends. We first use the MaxPooling operator, which is to select the maximum value in each column of a matrix to reduce the dimension, to generate embedding for the entire graph as:

$$\widetilde{\mathbf{z}}_t = \text{MaxPool}([\mathbf{z}_t^0, \mathbf{z}_t^1, \ldots, \mathbf{z}_t^N])$$

where $[\mathbf{z}_t^0, \mathbf{z}_t^1, \ldots, \mathbf{z}_t^N]$ is a matrix and the i-th column is $\mathbf{z}_t^i$, thus $\widetilde{\mathbf{z}}_t$ incorporates the most important features of all nodes extracted from the graph.

On the same day, the pandemic may have just emerged for some locations, but for other locations, the pandemic may have reached the peak. So we cannot model temporal patterns for all locations simultaneously using the same model parameters. We build a different model for each location. These locations share the same model structure and graph structure but have different model parameters. The STAN model is an end-to-end model, which means each location's model will adaptively extract most related spatio-temporal patterns from the attributed graph. All following equations are for one specific location, and we omit location index $i$ to reduce clutter.

We input the graph embedding to Gated Recurrent Unit (GRU)[17] network to learn temporal features. GRU is a type of recurrent neural networks, which can effectively model temporal sequences and is also widely used in many sequence analysis tasks[18, 19]. The GRU's hidden state is calculated as:

$$\mathbf{h}_t = GRU(\widetilde{\mathbf{z}_1}, \widetilde{\mathbf{z}_2}, \ldots, \widetilde{\mathbf{z}_t})$$

The obtained hidden state of the GRU $\mathbf{h}_t$ at the $t$-th timestep for a specific location contains both spatial and temporal patterns learned from real-world data.

## (IV). Multi-task Prediction and Transmission Dynamics Inspired

### Loss function

Our objective is to predict the number of infected cases for both long-term and short-term. In our method, we tackle this issue by using a multi-task learning framework to consider short-term and long-term prediction performance jointly.

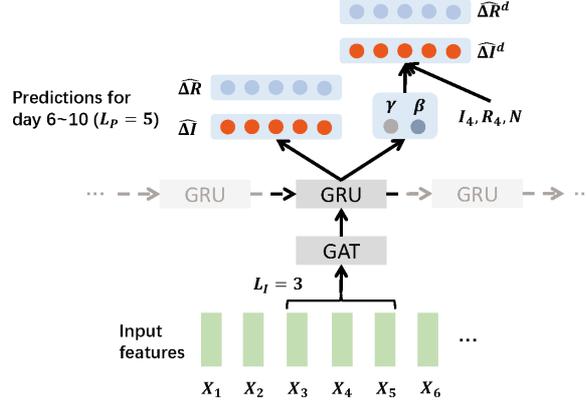

**Figure 2.** STAN prediction process for a single location at day 5 with $L_I = 3$ and $L_P = 5$.

The idea is to use short-term prediction loss and long-term transmission dynamics constraint loss to regularize learned hidden representations of node embeddings (i.e., hidden state of the GRU) $h_t$. In **Figure 2**, we provide the model prediction process for a single location on day 5 (the input window $L_I = 3$). Since the prediction process is the same for each timestep, we will omit the time index $t$ for simplicity. Concretely, the model output consists of two parts:

1. **Transmission/Recovery rate**. The traditional SIR-based model assumes that the disease transmission/recovery rate $\beta$ and $\gamma$ remain constant time. But in practice, those rates may easily change over time due to policies or disease evolution reasons. To solve this issue, we define a prediction window $L_P$ that the $\beta$ and $\gamma$ will be specific for this window. In **Figure 2**, $L_P = 5$ and the prediction window start from day 6 and end on day 10. The model will predict $\beta$ and $\gamma$ for the prediction window as:

$$\beta, \gamma = \text{sigmoid}(\text{MLP}(h))$$

where $\text{MLP}(\cdot)$ denotes the multi-layer perceptron, and we use sigmoid activation since both $\beta$ and $\gamma$ are between 0 and 1.

2. **Daily increased number of infected/recovered cases**. For the $i$-th prediction window, the model will predict the increment of the number of infected and recovered cases $\widehat{\Delta I}$ and $\widehat{\Delta R}$ as:

$$\widehat{\Delta I}, \widehat{\Delta R} = \text{MLP}(h)$$

Note that $\widehat{\Delta I}$ and $\widehat{\Delta R}$ are vectors since we are predicting for $L_P$ days.

To optimize the model parameters, we design two loss terms to make the model predict better for both short-term and long-term progression. Concretely, the loss function also consists of two parts:

1. **Transmission dynamics constraint loss**. The first loss term is a transmission dynamics constraint loss to regularize long-term prediction trends. In this term, we do not directly optimize the predictions using the ground truth numbers. Instead, we hope the model can use pandemic dynamics to regularize longer progressions. In particular, we use the predicted transmission/recovery rate $\beta$ and $\gamma$ to compute dynamic-based predictions $\widehat{\Delta I}^d$ and $\widehat{\Delta R}^d$. Then we can optimize the learned $\beta$ and $\gamma$ by optimizing these dynamic-based predictions,. Concretely, based on the SIR differential equations, give the prediction window in **Figure 2**, we can calculate the transmission dynamics-based increment number of infected cases and recovered cases iteratively as:

$$\widehat{\Delta I}^d = \left[\widehat{\Delta I}^d_{t+1}, \widehat{\Delta I}^d_{t+2}, \dots, \widehat{\Delta I}^d_{t+L_p}\right], each\ \widehat{\Delta I}^d_i = \beta S_{i-1} - \gamma I_{i-1} = \beta\left(N_P - \hat{I}^d_{i-1} - \hat{R}^d_{i-1}\right) - \gamma \hat{I}^d_{i-1}$$

$$\widehat{\Delta R}^d = \left[\widehat{\Delta R}^d_{t+1}, \widehat{\Delta R}^d_{t+2}, \dots, \widehat{\Delta R}^d_{t+L_p}\right], each\ \widehat{\Delta R}^d_i = \gamma \hat{I}^d_{i-1}$$

where $\hat{I}^d_{i-1}, \hat{R}^d_{i-1}$ can be calculated iteratively using the ground truth value of the infected and recovered cases at the day before the current prediction window (i.e., $I_4$ and $R_4$ in our example), $N_P$ denotes the population size of the current location and in Figure 2, $t = 5$ since we are making predictions at the 5-th time step. Finally, the transmission dynamics constraint loss is calculated as:

$$\mathcal{L}_d = \left(\widehat{\Delta I}^d - \Delta I\right)^2 + \left(\widehat{\Delta R}^d - \Delta R\right)^2$$

where $I_t$ and $R_t$ denote the ground truth number of infected and recovered cases. This loss term calculates the mean squared error of transmission dynamics-based predictions in order to make the prediction results in line with the long-term trend of pandemics.

2. **Prediction loss**. The second loss term is a regular mean squared error loss for the second task:

$$\mathcal{L}_r = \left(\widehat{\Delta I} - \Delta I\right)^2 + \left(\widehat{\Delta R} - \Delta R\right)^2$$

this loss term is to make the prediction results as close as possible to the short-term variation.

By combining two loss terms, the final loss function can be calculated by summing up all locations and timesteps as:

$$\mathcal{L} = \sum_i^T \sum_j^N (\mathcal{L}_r + \mathcal{L}_d)$$

## (V). Prediction with STAN

Once the model is trained, we can use the trained model to predict COVID cases at all locations given the length of input window $L_I$, the length of prediction window $L_p$, and graph information $G$ as input. Assuming our data is collected between $t_1$ and $t_2$, first, the graph information between $t_1$ and $t_2$ is fed into GAT to generate the graph embedding. Then the graph embeddings are fed into GRU. At the last time timestep of the GRU, the model will output predicted daily increased number of infected and recovered cases (i.e., $\widehat{\Delta I}, \widehat{\Delta R}$) for future $L_p$ days after $t_2$. Then we can calculate the total infected cases number by summing up the increased number. Note that we do not use the transmission dynamics constraints in the prediction time because this module is only used for optimizing the model in training time. We can obtain longer predictions by using a larger prediction window $L_P$ or adding the latest graph data incrementally.

# EXPERIMENTS

## Dataset Description

In this paper, we used a US county-level dataset that consists of COVID-19 related data from two resources: Johns Hopkins University (JHU) Coronavirus Resource Center[20] and IQVIA's claims data[21]. The data from JHU Coronavirus Resource Center was collected from March 22,

2020 to June 10, 2020. It has the number of active cases, confirmed cases, and deaths related to COVID-19 for different US locations. We select states with more than 1,000 confirmed cases by May 17 to ensure the data source accuracy, and finally we have 45 states and 193 counties in the dataset. For those counties, we set the number of cases before their respective first record dates as zero. The IQVIA's claims data is from the IQVIA US9 Database. We export patient claims data and prescription data from March 22, 2020 to June 10, 2020, from which we obtain the number of hospital and ICU visits and the term-frequency of each medical code per county per day. Detailed dataset descriptions are shown in Supplementary. The dataset has records for a total of 453,089 patients across the entire timespan of the JHU dataset. The 48 unique ICD-10 codes related to COVID-19 are listed in the Supplementary Material.

## Baseline Models

We compare STAN with the following baselines.
1. **SIR**: the susceptible-infected-removed (SIR), a basic disease transmission model that uses differential equations to simulate an epidemic. S, I, and R represent the number of susceptible, infected, and recovered individuals.
2. **SEIR**: the susceptible-exposed-infected-removed (SEIR) epidemiological model as another transmission dynamics constraint-based baselines. Compared to the SIR model, SEIR adds exposed population size to the equation.
3. **GRU**[17]: We input the latest number of infected cases into a naïve GRU and predict future numbers.
4. **ColaGNN**[9]: ColaGNN uses a location graph to extract spatial relationships for predicting pandemics. Different from STAN, graph nodes in ColaGNN only consist of time series of numbers of infected cases.
5. **CovidGNN**[10]: CovidGNN uses a graph neural network with skip connections to predict pandemics. They use the graph embedding to predict the future number of cases without using RNN to extract temporal relationships.

To explore the performance enhancement by transmission dynamics constraints and graph structures, we also compare STAN with the following reduced models.
1. **STAN-PC** removes transmission dynamics constraints from STAN.
2. **STAN-Graph** removes the GCN layers and graph data from STAN.

The implementation details of all models are shown in Supplementary Material.

## Tasks and Evaluation Strategy

We predict the future number of active cases on both county-level and state-level. To evaluate the ability of STAN for both long-term predictions and short-term predictions, we set the prediction window $L_P$ to 5, 15, and 20, i.e., predict for future 5, 15, and 20 days. All training sets start from March 22, and all test sets start from May 17. We also split $L_P$ days from the training sets as evaluation sets to determine model hyper-parameters. We set $L_I$ to 5. All locations are used in training and testing set by splitting along the time dimension, where early time windows are used for training the model, and later time windows are used for testing the model. All the models use the same training data and are also evaluated and tested using the same prediction time window.

We use the mean square error (MSE), mean absolute error (MAE) to evaluate our model. We also use the average concordance correlation coefficient (CCC) to evaluate the results. The CCC

measures the agreement between two variables, and it is computed as:

$$CCC = \frac{2\rho\sigma_x\sigma_y}{\sigma_x^2 + \sigma_y^2 + (\mu_x - \mu_y)^2}$$

where $\mu_x$ and $\mu_y$ are the means for the two variables, $\sigma_x^2$ and $\sigma_y^2$ are the corresponding variances. $\rho$ is the correlation coefficient between the two variables. Note that we chose not to use the coefficient of determination ($R^2$) is because the range of $R^2$ is $(-\infty, 1)$, so some extreme value may significantly affect the average value. But the range of CCC is between -1 and 1, so we can evaluate model results more reliably. To estimate a 95% confidence interval, according to previous pandemic research[22], we resample the locations 1,000 times, calculate the score on the resampled sets, and then use 2.5 and 97.5 percentiles of these scores as our confidence interval estimate.

# RESULTS

**Table 1** shows the average performance and a 95% confidence interval for state-level predictions of our model and all baseline models. STAN achieves the best performance under different lengths of the prediction window. When the length of the prediction window $L_P = 5$, STAN achieves 59% lower MSE, 33% lower MAE, and 23% higher CCC than the best baseline ColaGNN. When the length of the prediction window $L_P = 15$, STAN achieves 87% lower MSE, 56% lower MAE, and 47% higher CCC than ColaGNN. When the length of the prediction window $L_P = 20$, STAN achieves 48% lower MSE, 37% lower MAE, and 32% higher CCC than ColaGNN.

Table 1. Performance comparison for state-level predictions

| Model | MSE | MAE | CCC |
|---|---|---|---|
| **Prediction window $L_P = 5$** | | | |
| SIR | 2,968,711 | 921.06 | 0.41 |
| | (814,014-4,152,617) | (776.93-1209.22) | (0.37-0.45) |
| SEIR | 1,890,708 | 679.64 | 0.49 |
| | (612,049-3,562,890) | (681.57-1197.38) | (0.44-0.54) |
| GRU | 925,701 | 582.43 | 0.55 |
| | (501,309-1,792,855) | (479.50-842.38) | (0.50-0.60) |
| ColaGNN | 601,840 | 440.26 | 0.66 |
| | (381,907-982,354) | (323.57-568.44) | (0.59-0.72) |
| CovidGNN | 830,517 | 500.11 | 0.58 |
| | (430,127-1,109,311) | (367.55-645.72) | (0.53-0.64) |
| STAN-PC | 323,325 | 313.72 | 0.75 |
| | (213,702-450,314) | (280.39-392.01) | (0.70-0.79) |
| STAN-Graph | 472,245 | 362.04 | 0.67 |
| | (276,391-612,099) | (310.08-452.39) | (0.63-0.71) |
| **STAN** | **237,412** | **220.50** | **0.84** |
| | **(159,995-290,801)** | **(172.71-272.03)** | **(0.81-0.87)** |
| **Prediction window $L_P = 15$** | | | |
| Model | MSE | MAE | CCC |
| SIR | 22,939,910 | 2,438.95 | 0.32 |
| | (11,682,896-35,393,542) | (1,807.71-3,119.30) | (0.25-0.40) |

| Model | MSE | MAE | CCC |
|---|---|---|---|
| SEIR | 12,993,900 | 1,781.66 | 0.49 |
|  | (5,234,542-21,928,077) | (1,305.68-2,295.44) | (0.44-0.55) |
| GRU | 9,205,382 | 1,710.09 | 0.38 |
|  | (3,708,352-15,534,698) | (1,253.23-2,203.22) | (0.34-0.43) |
| ColaGNN | 7,192,031 | 1,290.41 | 0.57 |
|  | (2,897,281-12,137,035) | (945.67-1,662.52) | (0.51-0.63) |
| CovidGNN | 9,609,283 | 1,611.19 | 0.45 |
|  | (3,871,062-16,216,309) | (1,180.75-1,075.80) | (0.40-0.50) |
| STAN-PC | 1,785,304 | 774.22 | 0.72 |
|  | (1,032,754-2,895,702) | (650.39-904.74) | (0.68-0.76) |
| STAN-Graph | 2,897,053 | 964.09 | 0.66 |
|  | (1,352,076-4,309,806) | (784.01-1,011.36) | (0.60-0.71) |
| STAN | **972,192** | **586.56** | **0.84** |
|  | **(622,425-1,404,284)** | **(484.11-690.61)** | **(0.80-0.87)** |
| **Prediction window $L_P = 20$** | | | |
| Model | MSE | MAE | CCC |
| SIR | 46,732,397 | 3,439.25 | 0.25 |
|  | (23,701,239-71,720,863) | (2,538.28-4,423.32) | (0.18-0.33) |
| SEIR | 25,296,100 | 2,451.60 | 0.43 |
|  | (9,038,485-44,609,676) | (1,849.00-3,120.60) | (0.37-0.50) |
| GRU | 15,901,430 | 2,046.32 | 0.52 |
|  | (5,681,699-28,042,173) | (1,543.34-2,604.72) | (0.44-0.60) |
| ColaGNN | 9,317,132 | 1,645.42 | 0.63 |
|  | (3,971,773-19,667,264) | (1,240.98-2,094.72) | (0.54-0.72) |
| CovidGNN | 16,739,642 | 2,081.25 | 0.54 |
|  | (6,623,891-27,756,861) | (1,569.69-2,649.19) | (0.46-0.62) |
| STAN-PC | 5,929,321 | 1,209.41 | 0.75 |
|  | (3,082,515-10,309,710) | (1,032.75-1564.71) | (0.72-0.78) |
| STAN-Graph | 9,509,671 | 1,689.90 | 0.68 |
|  | (5,909,301-15,408,623) | (1342.09-2031.74) | (0.64-0.72) |
| STAN | **4,909,604** | **1,088.48** | **0.82** |
|  | **(1,999,607-8,811,535)** | **(820.78-1,366.01)** | **(0.79-0.86)** |

**Table 2** shows the performance for county-level prediction results. STAN also achieves the best performance under different lengths of the prediction window. When the length of the prediction window $L_P = 5$, STAN acquires 26% lower MSE, 29% lower MAE, and 25% higher CCC than ColaGNN. When the length of the prediction window $L_P = 15$, STAN achieves 55% lower MSE, 34% lower MAE, and 30% higher CCC than ColaGNN. When the length of the prediction window $L_P = 20$, STAN achieves 55% lower MSE, 37% lower MAE, and 29% higher CCC than ColaGNN.

Table 2. Performance comparison for county-level predictions

| **Prediction window $L_P = 5$** | | | |
|---|---|---|---|
| Model | MSE | MAE | CCC |

| Model | MSE | MAE | CCC |
|---|---|---|---|
| SIR | 93,512 | 151.33 | 0.40 |
| | (44,864-159,117) | (125.49-177.86) | (0.38-0.44) |
| SEIR | 134,494 | 165.14 | 0.35 |
| | (50,223-251,893) | (136.94-194.09) | (0.32-0.38) |
| GRU | 79,982 | 121.76 | 0.47 |
| | (39,820-136,096) | (100.96-143.10) | (0.43-0.51) |
| ColaGNN | 61,627 | 110.91 | 0.53 |
| | (36,176-104,864) | (91.97-130.36) | (0.49-0.58) |
| CovidGNN | 71,664 | 120.01 | 0.47 |
| | (37,718-121,941) | (99.16-140.55) | (0.43-0.51) |
| STAN-PC | 53,194 | 107.69 | 0.58 |
| | (32,961-103,211) | (87.63-123.12) | (0.53-0.63) |
| STAN-Graph | 50,331 | 104.99 | 0.57 |
| | (29,023-97,304) | (85.09-117.33) | (0.53-0.61) |
| STAN | **44,177** | **79.80** | **0.66** |
| | **(13,028- 79,916)** | **(66.17-93.79)** | **(0.60-0.71)** |

**Prediction window $L_P = 15$**

| Model | MSE | MAE | CCC |
|---|---|---|---|
| SIR | 884,249 | 415.79 | 0.29 |
| | (438,613-1,353,544) | (353.08-480.88) | (0.26-0.32) |
| SEIR | 1,102,601 | 391.06 | 0.33 |
| | (495,229-2,014,880) | (318.76-469.78) | (0.30-0.36) |
| GRU | 810,362 | 322.90 | 0.48 |
| | (382,092-1,008,423) | (291.54-397.06) | (0.45-0.51) |
| ColaGNN | 465,104 | 286.55 | 0.56 |
| | (290,623-878,780) | (258.72-352.37) | (0.54-0.59) |
| CovidGNN | 635,401 | 310.84 | 0.50 |
| | (342,312-935,801) | (280.64-382.23) | (0.47-0.53) |
| STAN-PC | 393,790 | 246.01 | 0.65 |
| | (251,323-675,304) | (215.42-305.33) | (0.63-0.68) |
| STAN-Graph | 339,082 | 242.74 | 0.66 |
| | (210,403-612,392) | (219.02-298.37) | (0.63-0.69) |
| STAN | **157,243** | **193.85** | **0.72** |
| | **(100,712-218,922)** | **(170.52-220.06)** | **(0.69-0.74)** |

**Prediction window $L_P = 20$**

| Model | MSE | MAE | CCC |
|---|---|---|---|
| SIR | 1,881,144 | 585.14 | 0.23 |
| | (511,054-2,953,878) | (497.63-682.66) | (0.20-0.26) |
| SEIR | 2,238,468 | 538.57 | 0.29 |
| | (647,839-3,998,162) | (437.36-644.32) | (0.26-0.32) |
| GRU | 981,064 | 461.86 | 0.46 |
| | (419,891-1,294,611) | (426.74-585.42) | (0.43-0.49) |
| ColaGNN | 703,377 | 410.13 | 0.55 |
| | (301,042-928,175) | (378.95-519.86) | (0.52-0.58) |
| CovidGNN | 1,043,261 | 468.44 | 0.43 |
| | (446,511-1,376,685) | (432.83-593.77) | (0.40-0.47) |
| STAN-PC | 492,374 | 271.63 | 0.67 |
| | (287,672-853,034) | (233.17-325.09) | (0.64-0.70) |
| STAN-Graph | 555,681 | 281.62 | 0.68 |
| | (300,626-879,030) | (239.01-356.71) | (0.64-0.71) |
| STAN | **326,258** | **253.86** | **0.71** |
| | **(187,532-505,796)** | **(219.53-291.86)** | **(0.69-0.74)** |

Table 3 T-Test for STAN and other baseline models (prediction window in 5 to 20 days)

| **Model** | State-5 | State-15 | State-20 | County-5 | County-15 | County-20 |
|---|---|---|---|---|---|---|
| SIR | 0.00E+00 | 0.00E+00 | 0.00E+00 | 0.00E+00 | 0.00E+00 | 0.00E+00 |
| SEIR | 0.00E+00 | 0.00E+00 | 4.04E-263 | 1.46E-203 | 6.21E-253 | 9.74E-257 |
| GRU | 5.27E-16 | 3.46E-12 | 2.70E-08 | 8.46E-18 | 3.57E-84 | 5.28E-56 |
| ColaGNN | 9.54E-09 | 4.08E-20 | 1.03E-04 | 2.83E-07 | 5.23E-60 | 9.85E-57 |
| CovidGNN | 2.02E-12 | 4.38E-24 | 5.27E-08 | 3.07E-20 | 4.32E-69 | 2.57E-61 |
| STAN-PC | 3.84E-04 | 6.23E-06 | 3.95E-03 | 1.98E-06 | 3.99E-32 | 7.21E-21 |
| STAN-Graph | 5.34E-07 | 6.62E-09 | 7.16E-05 | 4.76E-04 | 6.31E-16 | 4.55E-12 |

The results show STAN can conduct more accurate long-term and short-term prediction than SIR and SEIR models on both state-level and county-level. Since county-level graph data is more granular, STAN can benefit more by utilizing such data than the traditional dynamics-based model. It is also worth noting that both reduced model STAN-PC and STAN-Graph also outperform other baselines. This indicates that both transmission dynamics constraints and real-world evidence provide valuable information for pandemic progression prediction. We report the detailed performance of each location in the Supplementary Material. We conduct a T-test between STAN and each baseline model to check the performance difference statistically. The p-value is shown in **Table 3.** The results show that for each baseline model, STAN can significantly outperform statistically (p-value < 0.001). The detailed T-test results are shown in the Supplementary.

## DISCUSSION AND LIMITATIONS

In this section, we will discuss the advantages and also the limitations of our model. We draw the predicted curve of 20 days from May 16 to Jun 5 for two counties, El Paso, TX and Lake, IN, and two states, CA and MA. As shown in **Table 4**, for the two counties, STAN shows up to 99% relatively lower MSE compared to the SEIR and SIR model. For the two states, the performance improvement is much greater, STAN can achieve at most 95% lower MSE compared to the best SEIR models. And as shown in **Figure 3** and **Figure 4,** the curve also fits the actual trend better for both counties and states.

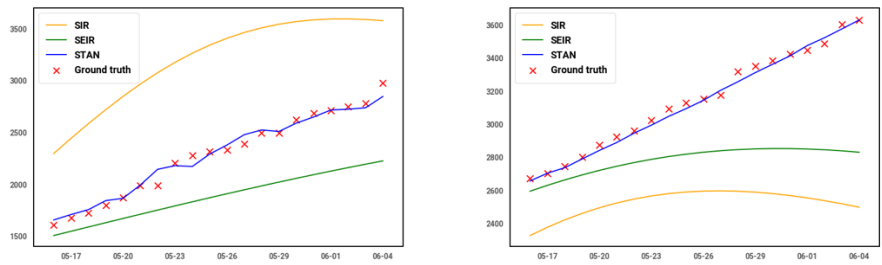

**Figure 3.** Predicted curve for two counties: El Paso, TX (left) and Lake, IN (right)

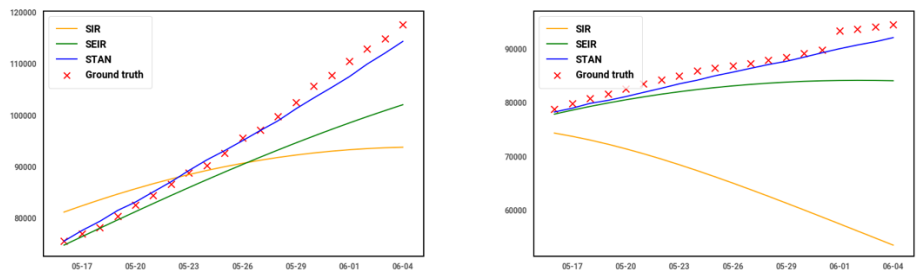

**Figure 4.** Predicted curve for two states: CA (left) and MA (right).

One obvious drawback of SIR and SEIR models is the overfitting issue. The SIR and SEIR models tend to predict the peak will come right after current data, which is especially apparent in the prediction curve of Lake, IN, and MA. This is because these traditional models do not incorporate the influence and interdependency of transmission between geographic regions. The characteristics of transmission of infectious diseases in one area are unlikely to be decoupled from those of nearby areas unless there are barriers to interaction between the regions such as topography (rivers with limited bridges or mountain ranges with limited road connections) or controlled borders. Such decoupling is infrequently present between counties in the USA. The inability to account for this geographic interdependency removes an important variable in the SIR and SEIR models and impedes their ability to predict the future progression using limited data at the early pandemic stage.

Table 4. Prediction performance for two counties (El Paso, TX and Lake, IN) and two states (CA and MA)

| Model | MSE | MAE | CCC |
|---|---|---|---|
| **El Paso, TX** | | | |
| SIR | 867,272 | 922.38 | 0.27 |
| SEIR | 196,954 | 403.36 | 0.47 |
| STAN | 3,889 | 47.14 | 0.99 |
| **Lake, IN** | | | |
| SIR | 448,839 | 619.16 | 0.06 |
| SEIR | 182,627 | 361.95 | 0.19 |
| STAN | 842 | 24.67 | 0.99 |
| **CA** | | | |
| SIR | 122,495,498 | 8,498.43 | 0.45 |
| SEIR | 58,379,735 | 5,848.03 | 0.79 |
| STAN | 2,782,908 | 1,306.15 | 0.99 |
| **MA** | | | |
| SIR | 592,535,941 | 21,652.34 | -0.11 |
| SEIR | 28,495,989 | 4,458.38 | 0.37 |
| STAN | 2,771,603 | 1,461.30 | 0.93 |

Though deep learning-based methods can achieve better performance than traditional statistical methods in various time series analysis and prediction tasks, there are still two major limitations in our work. The first limitation is the prediction window setting in our method. The traditional SIR and SEIR models use all historical training data to fit the model and generate the entire curve to be less affected by the fluctuations in the data. However, our model divides historical data into prediction windows in training time. This setting allows STAN to dynamically model the pandemic progressions. Therefore, it can better simulate the situations such as changes in reporting policy or dynamical changes in transmissibility or contagion. However, if the number of cases fluctuates drastically due to inaccuracy in the data collection process, it is difficult for the STAN model to learn valid and stable transmission and recovery rates. This issue can be further solved by applying dynamic data smoothing to smooth such abnormal data points.

The second limitation is that the transmission dynamics constraints may be too simple to reflect real-world situations such as home isolation and pandemic control policies. A lot of research focuses on improving the traditional SIR model by adding more population groups and transmission equations[3, 23]. These variants can be quickly adopted in our model by modifying the transmission dynamic loss.

The third limitation is about data quality for constructing the attribute graph. For node features, indeed, sometimes ICD codes may not reflect real pandemic status due to report delay or other reasons. And the edge mobility calculation can also be improved by incorporating more data sources such as traffic info or mobile geolocation tracking. Besides, the dynamic features and static features are processed together without considering their different characteristics. Future work can incorporate more data sources and extend the transmission dynamics constraints into STAN to further enhance the prediction performance with richer data and process different data types more reasonably. Our model can also be easily adopted for mortality or hospitalization prediction tasks by adding related statistics such as COVID-19 related treatment codes and ICU codes to the data.

## CONCLUSION

In this work, we propose a spatio-temporal attention network model (STAN) for the COVID-19 pandemic prediction. We map locations (e.g., a county or a state) to nodes on a graph. We use a set of static and dynamic features extracted from multiple real-world evidence, including real-world medical claims data, to construct nodes and use geographical proximity and demographic similarity between locations to construct edges. We use the graph attention network to incorporate the variant influence of the different neighboring locations of a node and predict the number of infected patients for a fixed period into the future. We also impose transmission dynamics constraints on predictions according to transmission dynamics. STAN achieves better prediction performance than the traditional SIR and SEIR model and other deep learning methods and shows less overfitting issues at the early stage of the pandemic. We hope our model can help governments and researchers better allocate medical resources and make policies to control the pandemic earlier. Our model can also be easily extended to predict hospitalization of COVID-19 as future work.

## DATA AVAILABILITY

The COVID-19 statistics data used in this article are provided by Johns Hopkins University (JHU) Coronavirus Resource Center[20], and they are publicly available at https://github.com/CSSEGISandData/COVID-19. The claims data used in this article are provided by IQVIA[21], and they will be shared on request to https://www.iqvia.com/solutions/real-world-evidence/real-world-data-and-insights.

## CONTRIBUTORS

Junyi Gao and Rakshith Sharma implemented the method and conducted the experiments. All authors were involved in developing the ideas and writing the paper.

## FUNDING

This work was in part supported by the National Science Foundation award SCH-2014438, IIS-1418511, CCF-1533768, IIS-1838042, the National Institute of Health award NIH R01 1R01NS107291-01 and R56HL138415.

## COMPETING INTERESTS

The authors have no competing interests to declare.

# SUPPLEMENTARY

# DATA DETAILS

The data from JHU Coronavirus Resource Center was collected from March 22, 2020 to June 10, 2020. It has the number of active cases, confirmed cases, and deaths related to COVID-19 for different US locations. We select states with more than 1,000 confirmed cases by May 17 to ensure the data source accuracy, and finally we have 45 states and 193 counties in the dataset. For those counties, we set the number of cases before their respective first record dates as zero.

The IQVIA's claims data is from the IQVIA US9 Database. The claims data are directly exported from the IQVIA's US claims database. This database contains adjudicated medical and pharmacy claims, including patient enrollment data for national and sub-national health plans and self-insured employer groups. We aggregated the hospitalization and ICU-related ICD codes to calculate the number of hospitalizations and ICU stays. The frequency of diagnosis codes is summed up for each location each day. The dataset has records for a total of 453,089 patients across the entire timespan of the JHU dataset. There are a total of 48 unique ICD-10 codes related to COVID-19 that were claimed from the set of codes considered, as shown in **Table 5**. Except the codes for defining COVID-19, we also include the codes of influenza and pneumonias because they are important symptoms of COVID-19. Even those features are not directly related to COVID-19, deep learning models are good at processing high-dimensional features and determine the proper weights assigning to those features. Hence, including those codes do not harm the model performance but may help the model to capture useful patterns from potential misdiagnosis and co-infections. Detailed county-level and state-level statistics are shown in **Table 6** and **Table 7**.

Table 5. ICD-10 (The 10th revision of the International statistical Classification of Diseases) codes used in our dataset. We choose codes that are relevant to the COVID-19 symptoms.

| ICD-10 | Description |
|---|---|
| R05 | Cough |
| R0602 | Shortness of breath |
| R509 | Fever, unspecified |
| U071 | COVID-19, virus identified |
| Z03818 | Encounter for observation for suspected exposure to other biological agents ruled out |
| Z20828 | Contact with and (suspected) exposure to other viral communicable diseases |
| B342 | Coronavirus infection, unspecified |
| B9729 | Other coronavirus as the cause of diseases classified elsewhere |
| J09 | Influenza due to certain identified influenza viruses |
| J10 | Influenza due to other identified influenza virus |
| J100 | Influenza due to other identified influenza virus with pneumonia |
| J101 | Influenza due to other identified influenza virus with other respiratory manifestations |
| J108 | Influenza due to other identified influenza virus with other manifestations |
| J11 | Influenza due to unidentified influenza virus |
| J110 | Influenza due to unidentified influenza virus with pneumonia |
| J111 | Influenza due to unidentified influenza virus with other respiratory manifestations |
| J118 | Influenza due to unidentified influenza virus with other manifestations |
| J12 | Viral pneumonia, not elsewhere classified |
| J120 | Adenoviral pneumonia |
| J1289 | Other viral pneumonia |
| J13 | Pneumonia due to Streptococcus pneumoniae |
| J14 | Pneumonia due to Hemophilus influenzae |
| J15 | Bacterial pneumonia, not elsewhere classified |
| J150 | Pneumonia due to Klebsiella pneumoniae |
| J151 | Pneumonia due to Pseudomonas |
| J152 | Pneumonia due to staphylococcus |
| J153 | Pneumonia due to streptococcus, group B |
| J154 | Pneumonia due to other streptococci |
| J155 | Pneumonia due to Escherichia coli |
| J156 | Pneumonia due to other Gram-negative bacteria |
| J157 | Pneumonia due to Mycoplasma pneumoniae |
| J158 | Pneumonia due to other specified bacteria |
| J159 | Unspecified bacterial pneumonia |
| J16 | Pneumonia due to other infectious organisms, not elsewhere classified |
| J160 | Chlamydial pneumonia |
| J168 | Pneumonia due to other specified infectious organisms |
| J17 | Pneumonia in diseases classified elsewhere |
| J18 | Pneumonia, unspecified organism |
| J180 | Bronchopneumonia, unspecified organism |
| J181 | Lobar pneumonia, unspecified organism |
| J182 | Hypostatic pneumonia, unspecified organism |
| J188 | Other pneumonia, unspecified organism |
| J189 | Pneumonia, unspecified organism |
| J208 | Acute bronchitis due to other specified organisms |
| J22 | Unspecified acute lower respiratory infection |
| J40 | Bronchitis, not specified as acute or chronic |
| J80 | Acute respiratory distress syndrome |
| J988 | Other specified respiratory disorders |

**Table 6.** County-level statistics

| County | Patient # | Bed # | Site # | Visit # | Avg code # | Male % | Avg male age | Avg female age |
|---|---|---|---|---|---|---|---|---|
| AL_JEFFERSON | 314758 | 4965 | 17 | 308 | 367 | 0.38 | 57 | 58 |
| AL_MOBILE | 136442 | 1983 | 8 | 274 | 369 | 0.42 | 61 | 63 |
| AZ_APACHE | 4494 | 160 | 4 | 2 | 4 | 0.41 | 51 | 52 |
| AZ_MARICOPA | 890970 | 11891 | 69 | 1385 | 1744 | 0.44 | 50 | 52 |
| AZ_NAVAJO | 14440 | 176 | 5 | 17 | 22 | 0.45 | 50 | 52 |
| AZ_PIMA | 176903 | 3566 | 20 | 187 | 234 | 0.44 | 53 | 53 |

| | | | | | | | |
|---|---|---|---|---|---|---|---|
| CA_ALAMEDA | 178694 | 3927 | 22 | 110 | 130 | 0.46 | 53 | 55 |
| CA_CONTRA COSTA | 202720 | 1739 | 9 | 167 | 197 | 0.41 | 56 | 58 |
| CA_FRESNO | 119063 | 3262 | 14 | 114 | 131 | 0.43 | 49 | 51 |
| CA_KERN | 161935 | 1890 | 15 | 155 | 184 | 0.4 | 53 | 57 |
| CA_LOS ANGELES | 1916739 | 28427 | 120 | 1006 | 1193 | 0.43 | 54 | 56 |
| CA_ORANGE | 825228 | 6490 | 34 | 589 | 671 | 0.45 | 57 | 58 |
| CA_RIVERSIDE | 270597 | 4284 | 25 | 153 | 181 | 0.43 | 59 | 60 |
| CA_SACRAMENTO | 380041 | 3274 | 17 | 172 | 202 | 0.42 | 53 | 55 |
| CA_SAN BERNARDINO | 268993 | 4967 | 30 | 123 | 142 | 0.42 | 56 | 57 |
| CA_SAN DIEGO | 408797 | 7997 | 32 | 263 | 318 | 0.41 | 56 | 58 |
| CA_SAN FRANCISCO | 76289 | 2676 | 13 | 47 | 56 | 0.41 | 53 | 55 |
| CA_SAN MATEO | 93222 | 2311 | 10 | 36 | 43 | 0.45 | 56 | 57 |
| CA_SANTA BARBARA | 69333 | 1151 | 8 | 40 | 48 | 0.43 | 54 | 57 |
| CA_SANTA CLARA | 175973 | 4831 | 17 | 88 | 104 | 0.43 | 48 | 47 |
| CA_TULARE | 98385 | 1502 | 6 | 60 | 67 | 0.42 | 48 | 49 |
| CO_ADAMS | 93915 | 1587 | 8 | 93 | 126 | 0.45 | 41 | 40 |
| CO_ARAPAHOE | 182502 | 1356 | 9 | 236 | 288 | 0.43 | 53 | 53 |
| CO_DENVER | 144436 | 2946 | 14 | 161 | 195 | 0.44 | 50 | 51 |
| CO_EL PASO | 83087 | 1833 | 11 | 70 | 87 | 0.45 | 46 | 41 |
| CO_JEFFERSON | 175778 | 748 | 5 | 238 | 289 | 0.46 | 55 | 54 |
| CO_WELD | 23798 | 598 | 5 | 17 | 23 | 0.45 | 51 | 50 |
| CT_FAIRFIELD | 187131 | 2428 | 9 | 214 | 282 | 0.43 | 53 | 53 |
| CT_HARTFORD | 226979 | 2271 | 13 | 241 | 320 | 0.43 | 57 | 59 |
| CT_LITCHFIELD | 28116 | 272 | 3 | 23 | 30 | 0.44 | 58 | 57 |
| CT_NEW HAVEN | 157756 | 3390 | 14 | 336 | 458 | 0.44 | 57 | 58 |
| DC_DISTRICT OF COLUMBIA | 226327 | 3510 | 12 | 260 | 336 | 0.41 | 47 | 49 |
| DE_KENT | 36709 | 400 | 3 | 47 | 58 | 0.41 | 48 | 52 |
| DE_NEW CASTLE | 193503 | 1409 | 10 | 176 | 234 | 0.44 | 50 | 49 |
| DE_SUSSEX | 60198 | 571 | 4 | 80 | 102 | 0.42 | 55 | 57 |
| FL_BROWARD | 455694 | 6711 | 26 | 654 | 825 | 0.44 | 53 | 53 |
| FL_COLLIER | 90050 | 1018 | 6 | 67 | 81 | 0.45 | 62 | 64 |
| FL_DUVAL | 341776 | 4198 | 15 | 275 | 340 | 0.44 | 54 | 55 |
| FL_HILLSBOROUGH | 685552 | 5022 | 22 | 451 | 542 | 0.43 | 59 | 61 |
| FL_LEE | 203463 | 1850 | 9 | 149 | 178 | 0.43 | 59 | 62 |
| FL_MIAMI-DADE | 693745 | 10704 | 40 | 812 | 1056 | 0.44 | 55 | 55 |
| FL_ORANGE | 358461 | 4627 | 23 | 429 | 528 | 0.43 | 51 | 52 |
| FL_PALM BEACH | 322627 | 4958 | 24 | 510 | 654 | 0.43 | 58 | 58 |

| | | | | | | | |
|---|---|---|---|---|---|---|---|
| FL_PINELLAS | 213141 | 4370 | 18 | 188 | 217 | 0.45 | 62 | 62 |
| GA_COBB | 140826 | 1442 | 5 | 173 | 221 | 0.41 | 54 | 56 |
| GA_DEKALB | 128609 | 2737 | 11 | 120 | 149 | 0.41 | 49 | 50 |
| GA_DOUGHERTY | 23680 | 698 | 2 | 17 | 25 | 0.38 | 48 | 55 |
| GA_FULTON | 292492 | 5165 | 16 | 317 | 406 | 0.41 | 52 | 55 |
| GA_GWINNETT | 162247 | 975 | 7 | 148 | 183 | 0.42 | 51 | 50 |
| GA_HALL | 65940 | 557 | 1 | 77 | 93 | 0.43 | 53 | 56 |
| IA_BLACK HAWK | 27536 | 671 | 3 | 29 | 36 | 0.44 | 53 | 56 |
| IA_POLK | 153287 | 2850 | 11 | 135 | 165 | 0.43 | 48 | 49 |
| IA_WOODBURY | 32998 | 584 | 2 | 44 | 52 | 0.45 | 47 | 49 |
| IL_COOK | 1047503 | 20995 | 80 | 1338 | 1852 | 0.43 | 49 | 51 |
| IL_DUPAGE | 326598 | 2454 | 11 | 403 | 581 | 0.43 | 54 | 55 |
| IL_KANE | 88889 | 1280 | 5 | 72 | 94 | 0.43 | 49 | 51 |
| IL_LAKE | 197634 | 1818 | 8 | 195 | 273 | 0.41 | 53 | 53 |
| IL_MCHENRY | 53933 | 468 | 4 | 63 | 86 | 0.44 | 49 | 50 |
| IL_WILL | 131192 | 1003 | 4 | 144 | 217 | 0.44 | 53 | 52 |
| IL_WINNEBAGO | 56610 | 1336 | 6 | 50 | 62 | 0.45 | 49 | 52 |
| IN_ALLEN | 154365 | 2062 | 12 | 173 | 256 | 0.43 | 55 | 58 |
| IN_CASS | 10605 | 268 | 3 | 16 | 27 | 0.42 | 53 | 55 |
| IN_LAKE | 90733 | 3102 | 15 | 221 | 321 | 0.44 | 55 | 57 |
| IN_MARION | 205269 | 5555 | 29 | 280 | 357 | 0.41 | 57 | 57 |
| IN_ST. JOSEPH | 69136 | 1045 | 8 | 81 | 106 | 0.43 | 54 | 56 |
| KS_FINNEY | 6000 | 102 | 1 | 1 | 1 | 0.45 | 46 | 44 |
| KS_FORD | 6740 | 99 | 1 | 11 | 14 | 0.43 | 49 | 52 |
| KS_LEAVENWORTH | 8386 | 429 | 3 | 6 | 7 | 0.47 | 60 | 61 |
| KS_WYANDOTTE | 21954 | 546 | 5 | 17 | 20 | 0.43 | 54 | 53 |
| KY_JEFFERSON | 283697 | 4707 | 17 | 364 | 464 | 0.42 | 52 | 54 |
| MA_BARNSTABLE | 46632 | 436 | 4 | 46 | 58 | 0.47 | 59 | 58 |
| MA_BRISTOL | 121026 | 1535 | 10 | 165 | 199 | 0.44 | 56 | 55 |
| MA_ESSEX | 193627 | 1645 | 14 | 273 | 376 | 0.41 | 55 | 56 |
| MA_HAMPDEN | 108738 | 1882 | 12 | 224 | 300 | 0.44 | 57 | 57 |
| MA_MIDDLESEX | 457272 | 4349 | 23 | 520 | 692 | 0.44 | 55 | 56 |
| MA_NORFOLK | 135880 | 1426 | 10 | 153 | 195 | 0.41 | 50 | 50 |
| MA_PLYMOUTH | 67009 | 1385 | 8 | 71 | 87 | 0.44 | 53 | 52 |
| MA_SUFFOLK | 188818 | 8339 | 24 | 323 | 487 | 0.44 | 52 | 53 |
| MA_WORCESTER | 113729 | 2313 | 16 | 163 | 203 | 0.45 | 55 | 54 |
| MD_ANNE ARUNDEL | 156853 | 629 | 2 | 179 | 234 | 0.43 | 50 | 53 |
| MD_BALTIMORE | 399555 | 6557 | 24 | 440 | 555 | 0.43 | 53 | 54 |
| MD_FREDERICK | 80759 | 269 | 1 | 80 | 116 | 0.45 | 53 | 55 |
| MD_HOWARD | 78631 | 576 | 3 | 100 | 127 | 0.42 | 51 | 53 |

|  |  |  |  |  |  |  |  |
|---|---|---|---|---|---|---|---|
| MD_MONTGOMERY | 278721 | 1776 | 10 | 366 | 466 | 0.42 | 48 | 50 |
| MD_PRINCE GEORGE'S | 212207 | 1057 | 7 | 233 | 298 | 0.41 | 52 | 53 |
| MI_GENESEE | 138912 | 1210 | 4 | 154 | 210 | 0.43 | 56 | 58 |
| MI_KENT | 365754 | 2319 | 12 | 269 | 327 | 0.42 | 50 | 53 |
| MI_MACOMB | 206173 | 1414 | 8 | 497 | 726 | 0.44 | 57 | 59 |
| MI_OAKLAND | 400174 | 4980 | 19 | 826 | 1190 | 0.44 | 56 | 57 |
| MI_WASHTENAW | 255637 | 2165 | 9 | 344 | 558 | 0.44 | 52 | 55 |
| MI_WAYNE | 415927 | 6963 | 30 | 828 | 1193 | 0.43 | 53 | 54 |
| MN_HENNEPIN | 451162 | 5751 | 17 | 337 | 451 | 0.43 | 51 | 54 |
| MN_NOBLES | 5990 | 48 | 1 | 6 | 9 | 0.47 | 51 | 54 |
| MN_RAMSEY | 205877 | 1885 | 7 | 83 | 100 | 0.41 | 49 | 51 |
| MN_STEARNS | 25176 | 569 | 4 | 7 | 8 | 0.46 | 53 | 58 |
| MO_ST. LOUIS | 314423 | 8401 | 34 | 249 | 317 | 0.42 | 55 | 56 |
| NC_DURHAM | 42238 | 1665 | 6 | 22 | 26 | 0.34 | 56 | 60 |
| NC_MECKLENBURG | 277203 | 3166 | 17 | 160 | 202 | 0.43 | 51 | 51 |
| NC_WAKE | 192235 | 2231 | 13 | 152 | 179 | 0.44 | 49 | 47 |
| ND_CASS | 24035 | 1300 | 9 | 15 | 19 | 0.45 | 49 | 51 |
| NE_DAKOTA | 4916 | 0 | 0 | 3 | 3 | 0.45 | 50 | 52 |
| NE_DOUGLAS | 133376 | 2995 | 19 | 97 | 122 | 0.43 | 51 | 53 |
| NE_HALL | 31979 | 159 | 1 | 44 | 61 | 0.42 | 51 | 53 |
| NH_HILLSBOROUGH | 81543 | 1110 | 8 | 88 | 116 | 0.44 | 52 | 53 |
| NH_ROCKINGHAM | 78225 | 601 | 6 | 126 | 152 | 0.44 | 56 | 57 |
| NJ_ATLANTIC | 57671 | 1567 | 6 | 62 | 77 | 0.44 | 54 | 55 |
| NJ_BERGEN | 570476 | 4410 | 14 | 833 | 1217 | 0.41 | 52 | 55 |
| NJ_BURLINGTON | 115186 | 1155 | 9 | 125 | 176 | 0.43 | 55 | 57 |
| NJ_CAMDEN | 172336 | 2625 | 9 | 277 | 383 | 0.43 | 56 | 55 |
| NJ_CUMBERLAND | 30538 | 362 | 3 | 42 | 55 | 0.45 | 53 | 54 |
| NJ_ESSEX | 265006 | 4210 | 13 | 456 | 653 | 0.41 | 52 | 53 |
| NJ_GLOUCESTER | 58628 | 515 | 2 | 53 | 67 | 0.42 | 49 | 51 |
| NJ_HUDSON | 86105 | 1812 | 7 | 250 | 354 | 0.42 | 53 | 53 |
| NJ_MERCER | 288422 | 1853 | 9 | 195 | 249 | 0.28 | 45 | 54 |
| NJ_MIDDLESEX | 203129 | 2704 | 11 | 607 | 852 | 0.43 | 54 | 55 |
| NJ_MONMOUTH | 168254 | 2681 | 8 | 481 | 783 | 0.44 | 53 | 53 |
| NJ_MORRIS | 149111 | 2113 | 10 | 408 | 662 | 0.45 | 53 | 54 |
| NJ_OCEAN | 136542 | 1707 | 9 | 213 | 271 | 0.47 | 54 | 53 |
| NJ_PASSAIC | 137475 | 1187 | 5 | 304 | 454 | 0.43 | 51 | 52 |
| NJ_SOMERSET | 291752 | 642 | 2 | 255 | 365 | 0.4 | 47 | 51 |
| NJ_UNION | 161568 | 2179 | 10 | 511 | 708 | 0.43 | 51 | 51 |
| NM_BERNALILLO | 111409 | 2838 | 18 | 107 | 129 | 0.42 | 54 | 55 |

| County | | | | | | | |
|---|---|---|---|---|---|---|---|
| NM_MCKINLEY | 3826 | 194 | 4 | 3 | 5 | 0.44 | 62 | 62 |
| NM_SAN JUAN | 17311 | 270 | 3 | 28 | 31 | 0.44 | 58 | 59 |
| NV_CLARK | 498324 | 6825 | 45 | 668 | 877 | 0.43 | 56 | 57 |
| NV_WASHOE | 83221 | 1965 | 13 | 100 | 130 | 0.44 | 54 | 55 |
| NY_ALBANY | 165355 | 1627 | 7 | 221 | 281 | 0.43 | 53 | 55 |
| NY_DUTCHESS | 118691 | 976 | 5 | 120 | 144 | 0.43 | 52 | 53 |
| NY_ERIE | 224088 | 3499 | 15 | 253 | 324 | 0.43 | 55 | 56 |
| NY_MONROE | 134568 | 1597 | 5 | 69 | 84 | 0.41 | 58 | 60 |
| NY_NASSAU | 670656 | 5045 | 16 | 1826 | 2749 | 0.44 | 51 | 52 |
| NY_NEW YORK | 1112742 | 10863 | 28 | 2450 | 3597 | 0.41 | 52 | 53 |
| NY_ONONDAGA | 147345 | 2031 | 7 | 125 | 149 | 0.42 | 54 | 56 |
| NY_ORANGE | 100906 | 1137 | 6 | 369 | 489 | 0.43 | 51 | 52 |
| NY_PUTNAM | 24692 | 347 | 2 | 34 | 44 | 0.4 | 54 | 54 |
| NY_ROCKLAND | 108328 | 1293 | 5 | 319 | 461 | 0.44 | 53 | 54 |
| NY_SUFFOLK | 379923 | 4718 | 16 | 1007 | 1468 | 0.43 | 51 | 52 |
| NY_SULLIVAN | 9473 | 174 | 2 | 14 | 19 | 0.44 | 52 | 54 |
| NY_ULSTER | 41779 | 175 | 2 | 63 | 83 | 0.44 | 57 | 57 |
| NY_WESTCHESTER | 339702 | 4401 | 21 | 755 | 1062 | 0.42 | 52 | 53 |
| OH_CUYAHOGA | 373592 | 7501 | 32 | 365 | 454 | 0.42 | 54 | 56 |
| OH_FRANKLIN | 570600 | 6918 | 31 | 376 | 459 | 0.41 | 47 | 49 |
| OH_HAMILTON | 415904 | 4922 | 23 | 382 | 478 | 0.43 | 46 | 45 |
| OH_LUCAS | 213970 | 3351 | 17 | 198 | 269 | 0.42 | 52 | 54 |
| OH_MAHONING | 70179 | 1147 | 9 | 170 | 225 | 0.43 | 55 | 56 |
| OH_MARION | 21146 | 270 | 1 | 15 | 18 | 0.42 | 49 | 52 |
| OH_PICKAWAY | 11713 | 94 | 1 | 5 | 6 | 0.42 | 53 | 56 |
| OK_OKLAHOMA | 202583 | 5371 | 42 | 181 | 220 | 0.42 | 56 | 57 |
| PA_ALLEGHENY | 356957 | 7765 | 34 | 295 | 351 | 0.43 | 57 | 58 |
| PA_BERKS | 75509 | 1399 | 7 | 85 | 114 | 0.43 | 54 | 58 |
| PA_BUCKS | 208543 | 1343 | 9 | 244 | 345 | 0.42 | 56 | 57 |
| PA_CHESTER | 181856 | 1011 | 7 | 134 | 168 | 0.42 | 53 | 54 |
| PA_DELAWARE | 148302 | 1668 | 9 | 143 | 192 | 0.42 | 54 | 54 |
| PA_LACKAWANNA | 52668 | 1042 | 6 | 68 | 86 | 0.44 | 58 | 59 |
| PA_LANCASTER | 100228 | 1093 | 6 | 71 | 96 | 0.43 | 50 | 51 |
| PA_LEHIGH | 111238 | 1372 | 9 | 182 | 246 | 0.4 | 50 | 52 |
| PA_LUZERNE | 61895 | 501 | 6 | 50 | 62 | 0.42 | 58 | 59 |
| PA_MONROE | 30662 | 345 | 2 | 45 | 62 | 0.42 | 51 | 53 |
| PA_MONTGOMERY | 418518 | 3207 | 16 | 339 | 458 | 0.4 | 57 | 59 |
| PA_NORTHAMPTON | 91598 | 1171 | 6 | 130 | 177 | 0.43 | 49 | 52 |

| | | | | | | | |
|---|---|---|---|---|---|---|---|
| PA_PHILADELPHIA | 344652 | 8747 | 36 | 629 | 891 | 0.43 | 49 | 51 |
| RI_PROVIDENCE | 251120 | 3010 | 13 | 358 | 424 | 0.43 | 52 | 53 |
| SC_GREENVILLE | 170969 | 1793 | 15 | 138 | 175 | 0.43 | 51 | 51 |
| SC_RICHLAND | 138916 | 2545 | 13 | 95 | 113 | 0.42 | 50 | 50 |
| SD_MINNEHAHA | 78776 | 1487 | 10 | 137 | 170 | 0.43 | 56 | 59 |
| TN_DAVIDSON | 375090 | 3570 | 16 | 208 | 257 | 0.4 | 51 | 54 |
| TN_SHELBY | 319585 | 5388 | 26 | 474 | 620 | 0.4 | 52 | 55 |
| TN_TROUSDALE | 568 | 25 | 1 | 0 | 0 | 0.41 | 58 | 53 |
| TX_BEXAR | 328337 | 8432 | 46 | 321 | 367 | 0.43 | 54 | 54 |
| TX_COLLIN | 302636 | 2446 | 21 | 237 | 290 | 0.4 | 53 | 53 |
| TX_DALLAS | 716189 | 9899 | 56 | 753 | 953 | 0.38 | 51 | 54 |
| TX_DENTON | 173703 | 1638 | 19 | 144 | 179 | 0.43 | 52 | 53 |
| TX_EL PASO | 97053 | 3043 | 20 | 123 | 140 | 0.44 | 56 | 56 |
| TX_FORT BEND | 97058 | 1380 | 13 | 138 | 180 | 0.42 | 51 | 53 |
| TX_HARRIS | 758706 | 17394 | 95 | 995 | 1264 | 0.43 | 47 | 48 |
| TX_POTTER | 68856 | 1403 | 9 | 146 | 172 | 0.43 | 57 | 58 |
| TX_TARRANT | 361685 | 6122 | 44 | 514 | 656 | 0.42 | 53 | 55 |
| TX_TRAVIS | 694270 | 3367 | 23 | 633 | 801 | 0.39 | 49 | 52 |
| UT_SALT LAKE | 162528 | 2451 | 19 | 143 | 176 | 0.43 | 51 | 51 |
| UT_UTAH | 63606 | 1353 | 10 | 44 | 55 | 0.45 | 47 | 46 |
| VA_ARLINGTON | 55046 | 370 | 1 | 63 | 89 | 0.41 | 49 | 51 |
| VA_CHESTERFIELD | 97216 | 490 | 3 | 44 | 51 | 0.42 | 51 | 51 |
| VA_FAIRFAX | 368091 | 2013 | 8 | 335 | 455 | 0.42 | 49 | 49 |
| VA_HENRICO | 194873 | 704 | 4 | 155 | 183 | 0.42 | 50 | 47 |
| VA_LOUDOUN | 95930 | 487 | 5 | 119 | 156 | 0.44 | 49 | 49 |
| VA_PRINCE WILLIAM | 62217 | 243 | 2 | 62 | 87 | 0.42 | 48 | 49 |
| WA_KING | 369636 | 5871 | 28 | 341 | 440 | 0.42 | 50 | 52 |
| WA_PIERCE | 172742 | 2541 | 11 | 182 | 231 | 0.43 | 48 | 49 |
| WA_SNOHOMISH | 84202 | 1017 | 8 | 84 | 111 | 0.43 | 51 | 52 |
| WA_YAKIMA | 36485 | 528 | 4 | 52 | 64 | 0.44 | 53 | 55 |
| WI_BROWN | 75291 | 1278 | 8 | 73 | 94 | 0.45 | 53 | 57 |
| WI_MILWAUKEE | 179429 | 5441 | 22 | 281 | 347 | 0.42 | 50 | 53 |
| WI_RACINE | 17972 | 665 | 4 | 5 | 6 | 0.43 | 48 | 48 |

**Table 7.** State-level statistics

| State | Patient # | Bed # | Site # | Visit # | Avg code # | Male % | Avg male age | Avg female age |
|---|---|---|---|---|---|---|---|---|
| CA | 11280624 | 95697 | 510 | 3679 | 4305 | 0.43 | 55.43 | 56.72 |
| NJ | 6318697 | 32850 | 134 | 5258 | 7573 | 0.43 | 52.67 | 54.1 |
| NY | 11672957 | 66807 | 266 | 11791 | 16587 | 0.43 | 54.77 | 56.34 |
| WV | 864456 | 8527 | 75 | 402 | 474 | 0.43 | 56.33 | 57.11 |

|  |  |  |  |  |  |  |  |
|---|---:|---:|---:|---:|---:|---:|---:|
| VA | 4086215 | 12590 | 86 | 1205 | 1519 | 0.42 | 54.59 | 54.75 |
| MN | 2230003 | 17462 | 157 | 681 | 863 | 0.44 | 56.59 | 58.03 |
| CT | 1371913 | 10469 | 50 | 922 | 1226 | 0.44 | 56.75 | 57.5 |
| GA | 4672277 | 30709 | 196 | 2341 | 2922 | 0.4 | 56.01 | 56.44 |
| WI | 2163502 | 19661 | 160 | 785 | 994 | 0.44 | 54.44 | 56.93 |
| IA | 1541463 | 12487 | 130 | 661 | 844 | 0.44 | 52.27 | 53.39 |
| AR | 1142575 | 12755 | 111 | 642 | 779 | 0.44 | 58 | 57.93 |
| OK | 1505373 | 16185 | 175 | 804 | 997 | 0.43 | 59.47 | 59.69 |
| NC | 4393806 | 29450 | 180 | 1586 | 1905 | 0.42 | 56.8 | 56.56 |
| OR | 1736670 | 10202 | 77 | 615 | 779 | 0.43 | 55.19 | 56.89 |
| TX | 12470003 | 93267 | 719 | 6318 | 7796 | 0.43 | 55.52 | 56.05 |
| ND | 286099 | 3782 | 57 | 118 | 148 | 0.44 | 52.33 | 53.25 |
| OH | 7160485 | 46531 | 282 | 3230 | 4012 | 0.43 | 52.28 | 53.92 |
| SD | 468914 | 4499 | 72 | 252 | 310 | 0.44 | 59.11 | 59.95 |
| ID | 603338 | 4199 | 60 | 263 | 318 | 0.44 | 55.42 | 57.33 |
| CO | 2340901 | 15125 | 124 | 1183 | 1474 | 0.44 | 53.48 | 53.89 |
| NE | 848369 | 7791 | 107 | 344 | 439 | 0.45 | 53.93 | 55.49 |
| KY | 2869138 | 18306 | 130 | 1474 | 1828 | 0.42 | 51.74 | 51.52 |
| FL | 10673839 | 71431 | 364 | 5620 | 6899 | 0.44 | 59.91 | 60.88 |
| KS | 1245960 | 13062 | 172 | 445 | 553 | 0.44 | 53.5 | 54.22 |
| UT | 918481 | 5914 | 64 | 341 | 419 | 0.46 | 50.79 | 50.93 |
| AL | 2307017 | 20490 | 132 | 1534 | 1924 | 0.43 | 59.61 | 59.49 |
| ME | 749670 | 4533 | 43 | 336 | 416 | 0.43 | 57.69 | 60.81 |
| IL | 5313832 | 39999 | 229 | 2887 | 3966 | 0.43 | 52.61 | 53.69 |
| NM | 605791 | 5587 | 58 | 217 | 255 | 0.43 | 52.73 | 53.39 |
| PA | 6351803 | 46230 | 281 | 3273 | 4297 | 0.43 | 56.01 | 57.15 |
| TN | 4199028 | 25133 | 169 | 2021 | 2504 | 0.43 | 56.91 | 56.53 |
| WA | 2661115 | 17129 | 127 | 1222 | 1558 | 0.43 | 48.9 | 50.92 |
| IN | 3055544 | 23641 | 198 | 1536 | 2034 | 0.43 | 56.92 | 57.8 |
| MO | 1932287 | 23007 | 164 | 850 | 1064 | 0.43 | 56.96 | 57.39 |
| MI | 5543648 | 30763 | 190 | 3764 | 5276 | 0.44 | 56.27 | 58.5 |
| SC | 2277366 | 14817 | 104 | 1144 | 1424 | 0.41 | 53.91 | 54.46 |
| NH | 547482 | 3101 | 33 | 306 | 385 | 0.44 | 56.5 | 58.2 |
| MS | 1247700 | 16605 | 130 | 749 | 925 | 0.41 | 56.07 | 55.89 |
| AZ | 2987234 | 18730 | 133 | 1847 | 2318 | 0.43 | 51.33 | 52.33 |
| MA | 2978862 | 24108 | 130 | 2018 | 2698 | 0.44 | 53.57 | 54.43 |
| NV | 1178120 | 9380 | 73 | 817 | 1069 | 0.46 | 55 | 56.38 |
| RI | 783321 | 3723 | 17 | 552 | 659 | 0.44 | 53.6 | 53.2 |
| MD | 3541594 | 13848 | 71 | 1908 | 2434 | 0.43 | 51.48 | 53.43 |
| DE | 577730 | 2380 | 17 | 303 | 395 | 0.42 | 51 | 52.67 |
| DC | 383908 | 3510 | 12 | 260 | 336 | 0.41 | 47 | 49 |

# IMPLEMENTATION DETAILS

All methods are implemented in PyTorch 1.1 and trained on a server equipped with an Intel Xeon E5-2620 Octa-Core CPU, 256GB Memory and a Titan V GPU. For the hyper-parameters of baseline models, we follow the recommended setting if it is available in the original paper. Otherwise, we determine its value by grid search on the validation set.

For the STAN model, the hidden dimension of the GRU is set to 200 and the hidden dimension of the MLP is set to 100. The graph embedding dimension is set to 400 and the graph attention dimension is set to 650. The input sliding window $L_I$ is set to 5. The $r$ is set to 30, $\alpha$ and $\beta$ is set to 0.35 and 0.37.

For the GRU model, the hidden dimension is set to 100. For the ColaGNN, the hidden dimension of GRU is set to 256, the dimension of graph node embedding is set to 500. For the CovidGNN model, we use a two-layer GNN and the dimension of graph node embedding is set to 256.

# PERFORMANCE DETAILS

We report the prediction MSE for each state and county in **Table 8** and **Table 9**. Due to space limits, we only take SIR and SEIR into comparison. For all 45 states, when $L_P = 5$, STAN achieves the best performance on 37 states; when $L_P = 15$ and 20, STAN achieves the best performance on 35 states. For all 45 states, when $L_P = 5$, STAN achieves the best performance on 37 states; when $L_P = 15$ and 20, STAN achieves the best performance on 35 states.

For all 193 counties, when $L_P = 5$, STAN achieves the best performance on 134 counties; when $L_P = 15$, STAN achieves the best performance on 148 states; when $L_P = 20$, STAN achieves the best performance on 143 states.

While conducting county-level prediction, STAN can achieve better long-term prediction compared to SIR and SEIR on most locations. This is due to the location graph is more granular and the model can extract detailed spatial interactions between nodes. And also for some locations, the pandemic haven't outbreak, so STAN can better predict future progression by considering progressions from neighboring locations. While aggregating the data and conducting state-level predictions, STAN's performance is more consistent over all length of prediction window.

**Table 8.** State-wise prediction MSE

| State | $L_P = 5$ | | | $L_P = 15$ | | | $L_P = 20$ | | |
|---|---|---|---|---|---|---|---|---|---|
| | SIR | SEIR | STAN | SIR | SEIR | STAN | SIR | SEIR | STAN |
| AL | 35969 | **33840** | 50841 | **1148638** | 1325467 | 2143502 | **2245236** | 2717692 | 2934974 |
| AR | 525720 | 284850 | **17204** | 3725935 | 1829427 | **863864** | 7632371 | 3919630 | **1223407** |
| AZ | 47235 | 116290 | **43536** | 213314 | 771706 | **29750** | 2036012 | 4029891 | **248002** |
| CA | 2822766 | **638352** | 1781510 | 58355218 | 19821426 | **788942** | 156482909 | 58379735 | **2782909** |
| CO | 573072 | 130782 | **829** | 6361723 | 2275589 | **12394** | 14264220 | 5944874 | **422423** |
| CT | 5081367 | 1917780 | **141097** | 19189818 | 3110738 | **717895** | 33225807 | 4117263 | **1538260** |
| DC | **3221** | 31441 | 18307 | **12073** | 507313 | 61560 | 44444 | 1060394 | **31670** |
| DE | 216513 | 19855 | **5568** | 1173115 | 11306 | 36567 | 2134180 | **25150** | 84301 |
| FL | 1333562 | 770617 | **156373** | 8958648 | 3979007 | **3735569** | 19716893 | 8349910 | **5438116** |
| GA | 478806 | **40364** | 188259 | 5150993 | 8394772 | **2572506** | 7531756 | 6422158 | **529251** |
| IA | 790491 | 341599 | **9753** | 11584529 | 4250450 | **499438** | 23847090 | 8362259 | **1243471** |
| ID | 3968 | **2679** | 2709 | 24947 | **13793** | 14486 | 53712 | **28624** | 30156 |
| IL | 7677558 | 4539416 | **787052** | 87050468 | 47549378 | **11871478** | 174425565 | **90124497** | 102453656 |
| IN | 998649 | 1207340 | **14134** | 9804565 | 9782019 | **77851** | 22376904 | 20784999 | **99796** |
| KS | 105276 | 31842 | **25305** | 1856691 | 771430 | **131930** | 4089669 | 1838994 | **712897** |
| KY | 49469 | 50893 | **9794** | 275923 | 253590 | **174215** | 987554 | 886238 | **40289** |
| MA | 34638912 | 3430794 | **49765** | 188760053 | 10321814 | **694645** | 376509655 | 28495989 | **2771604** |
| MD | 2783292 | 3147157 | **82780** | 5010187 | 4771543 | **840472** | 17082240 | 15415005 | **1418938** |
| ME | 6411 | **1054** | 3201 | 40369 | **17499** | 28308 | 58277 | **27665** | 68965 |
| MI | 5163810 | 1825336 | **29887** | 4247109 | 1762722 | **228764** | 12098923 | 1337059 | **661681** |
| MN | 88407 | 1456259 | **2704** | 3197951 | 6891496 | 4023834 | **8432004** | 10666408 | 18165335 |
| MO | 23671 | 132373 | **1097** | 4317008 | 2754554 | **4925** | 3442633 | 2130238 | **7627** |
| MS | 98457 | 83079 | **21409** | 1661032 | 1445189 | **133923** | 4058397 | 3568215 | **80754** |
| NC | 642968 | 388320 | **68220** | 8984167 | 4927417 | **4438426** | 21713680 | 11804975 | **7397819** |
| ND | 33225 | 10393 | **7156** | 223308 | 48529 | **36832** | 333595 | 45467 | **41254** |
| NE | 1797250 | 452137 | **15271** | 12913421 | 2482806 | **134415** | 24967634 | 5231907 | **1761118** |
| NH | 7540 | 14549 | **5795** | 122793 | 187446 | **10922** | 286786 | 405453 | **15569** |
| NJ | 21547538 | 11697626 | **1364570** | 165123676 | 102546164 | **430633** | 328017796 | 218066841 | **1153618** |
| NM | 101140 | 20694 | **2188** | 248806 | 39101 | **2572** | 742942 | 78696 | **56863** |
| NV | 114262 | 47199 | **27649** | 768725 | 234978 | **126553** | 1490263 | 409414 | **5774** |
| NY | 41029629 | 32073349 | **1439871** | 344142511 | 276121074 | **265175** | 705405848 | 572633149 | **62852285** |
| OH | 99736 | 70303 | **33923** | 154392 | 2612732 | 239138 | **212192** | 6303436 | 704602 |
| OK | 18380 | 24159 | **1264** | 80102 | 118184 | **13805** | 134715 | 205189 | **26168** |
| OR | 1604 | 2754 | **603** | 29013 | 28568 | **23733** | 37285 | 32196 | **10637** |
| PA | 3487833 | 342445 | **274261** | 32691226 | 4210910 | **976499** | 65992492 | **9148289** | 10026111 |
| RI | 626385 | **59254** | 116616 | 1940867 | **497888** | 1862138 | 3743923 | **373901** | 2130920 |
| SC | 205482 | 70740 | **5185** | 1888462 | 555102 | **157086** | 4486233 | 1418521 | **484925** |
| SD | 28376 | 21351 | **1600** | 105991 | 59664 | **32060** | 170071 | 83093 | **2859** |
| TN | 556642 | 549156 | **90940** | 6200060 | 5601928 | **1392780** | 15920345 | 14173437 | **3181705** |
| TX | 861329 | 849580 | **105879** | 1025252 | **836880** | 2030543 | 5769692 | 4537729 | **980812** |
| UT | 72980 | 42764 | **3107** | 141176 | 95690 | **73101** | 626564 | 344090 | **161141** |
| VA | 1496684 | 519132 | **97442** | 19820177 | 32076705 | **1532963** | 15491286 | 26336937 | **3347044** |
| WA | **10767** | 69474 | 107561 | **51859** | 840258 | 356025 | **300588** | 2743506 | 1450136 |
| WI | 242847 | 420722 | **18915** | 4281439 | 5991377 | **166584** | 9248112 | 12308700 | **499113** |
| WV | 1435 | 513 | **291** | 55758 | 31789 | **24997** | 103921 | 55455 | **13567** |

**Table 9.** County-wise prediction MSE

| County | $L_P = 5$ | | | $L_P = 15$ | | | $L_P = 20$ | | |
|---|---|---|---|---|---|---|---|---|---|
| | SIR | SEIR | STAN | SIR | SEIR | STAN | SIR | SEIR | STAN |
| AL_JEFFERSON | 3223 | 3574 | **2361** | 25026 | 23800 | **20589** | 42780 | 38176 | **36682** |
| AL_MOBILE | 5752 | 2438 | **505** | 80077 | 38524 | **660** | 157201 | 75622 | **746** |
| AZ_APACHE | 450 | 1029 | **442** | 1649 | 3421 | **864** | 1628 | 9840 | **859** |

| County | | | | | | | | |
|---|---|---|---|---|---|---|---|---|
| AZ_MARICOPA | 85114 | 65174 | **6480** | 674175 | 402777 | **21385** | 2582965 | 1686594 | **51516** |
| AZ_NAVAJO | 2640 | 4399 | **2606** | **11593** | 20067 | 34941 | **25471** | 41677 | 76257 |
| AZ_PIMA | 845 | 1173 | **282** | 4314 | 4797 | **608** | 29433 | 28692 | **8965** |
| CA_ALAMEDA | **740** | 1299 | 18402 | 33418 | **30643** | 36888 | 93111 | 78405 | **42128** |
| CA_CONTRA COSTA | 2802 | 3619 | **1731** | 28556 | 36088 | **11313** | 55478 | 69388 | **17466** |
| CA_FRESNO | **3067** | 5511 | 17816 | **2465** | 14245 | 51646 | **4838** | 32615 | 90271 |
| CA_KERN | **228** | 365 | 1702 | 9908 | 15342 | 21013 | **40671** | 57632 | 56686 |
| CA_LOS ANGELES | 887585 | 532537 | **72756** | 20882437 | 12642913 | **5977146** | 55618021 | 33625707 | **14844293** |
| CA_ORANGE | 9210 | 16150 | **3056** | 33995 | 122615 | 119575 | **78495** | 296715 | 230079 |
| CA_RIVERSIDE | 14398 | 19619 | **7958** | 186497 | 196933 | **106008** | 427835 | 402958 | **107403** |
| CA_SACRAMENTO | 1210 | 878 | **11** | 9750 | 6244 | **294** | 24057 | 15512 | **698** |
| CA_SAN BERNARDINO | **458** | 4293 | 4789 | **52753** | 169233 | 212855 | **112603** | 369057 | 312516 |
| CA_SAN DIEGO | 27754 | **8096** | 277704 | 282128 | **37043** | 936850 | 561704 | **42858** | 1381216 |
| CA_SAN FRANCISCO | 8611 | **611** | 9017 | 151919 | 32332 | **10709** | 294164 | 64677 | **9484** |
| CA_SAN MATEO | 2079 | 3696 | **424** | 47068 | 67371 | **9153** | 96857 | 135225 | **19984** |
| CA_SANTA BARBARA | 88832 | 17003 | **5688** | 357411 | 8953 | **8620** | 438890 | **8125** | 9160 |
| CA_SANTA CLARA | **2074** | 2707 | 3711 | 31383 | 33590 | **14573** | 56619 | 58146 | **23667** |
| CA_TULARE | 3463 | **3064** | 22501 | 68854 | **14434** | 53733 | 174311 | **23965** | 61017 |
| CO_ADAMS | 1083 | **588** | 961 | 58570 | 41413 | **1480** | 147363 | 104163 | **2291** |
| CO_ARAPAHOE | 10301 | **1800** | 6420 | 190753 | 61265 | **9943** | 406163 | 135604 | **22915** |
| CO_DENVER | 48841 | 25151 | **18333** | 731529 | 468192 | **21210** | 1375704 | 898040 | **16347** |
| CO_EL PASO | 4793 | **4787** | 12658 | 41470 | 41404 | **13538** | 51033 | 50575 | **13188** |
| CO_JEFFERSON | **107** | 752 | 3291 | **6154** | 13605 | 17224 | **8887** | 18918 | 23809 |
| CO_WELD | 209 | 2651 | **203** | 13406 | 42649 | **4979** | 32513 | 87028 | **13056** |
| CT_FAIRFIELD | **5637** | 239179 | 75862 | **51560** | 1032925 | 157826 | **40873** | 1475549 | 399440 |
| CT_HARTFORD | 208727 | **164560** | 363294 | 413916 | 326912 | **256489** | 340851 | 256428 | **206795** |
| CT_LITCHFIELD | **1080** | 3299 | 1635 | 2234 | 11614 | **1201** | **2089** | 16346 | 2888 |
| CT_NEW HAVEN | **21125** | 32971 | 44292 | 59218 | 116076 | **59152** | 49581 | 110474 | 134604 |
| DC_DISTRICT OF COLUMBIA | 42435 | **31369** | 84285 | 790182 | 642361 | **62862** | 1554373 | 1226894 | **46607** |
| DE_KENT | **66** | 448 | 70 | **400** | 1038 | 804 | 1137 | **1007** | 6249 |
| DE_NEW CASTLE | **936** | 957 | 11428 | **62516** | 65012 | 63635 | **76431** | 80919 | 78401 |
| DE_SUSSEX | **1951** | 16321 | 26268 | 93905 | 26174 | **18834** | 211874 | 49672 | **43961** |
| FL_BROWARD | 14454 | **5659** | 7503 | 229337 | 124927 | **41251** | 514592 | 297255 | **33140** |
| FL_COLLIER | **3511** | 4014 | 4995 | **50245** | 55995 | 53655 | **99613** | 112278 | 111620 |
| FL_DUVAL | 1528 | 1149 | **79** | 11303 | 8210 | **1107** | 22112 | 15954 | **4285** |
| FL_HILLSBOROUGH | 7063 | 3316 | **152** | 107204 | 57060 | **2537** | 260544 | 145734 | **15821** |
| FL_LEE | 3127 | 4147 | **980** | 29839 | 33191 | **1494** | 60561 | 63496 | **4864** |
| FL_MIAMI-DADE | 308595 | 284310 | **12765** | 915513 | 658544 | **562077** | 1553852 | 977557 | **802164** |
| FL_ORANGE | 5476 | 4007 | **3212** | 25217 | 19498 | **474** | 51890 | 41489 | **2855** |
| FL_PALM BEACH | 854 | **786** | 5374 | **3656** | 10391 | 25501 | 33870 | 58584 | **26682** |
| FL_PINELLAS | 3448 | 1982 | **31** | 9357 | 4491 | **229** | 15304 | 7335 | **521** |

| County | | | | | | | | |
|---|---|---|---|---|---|---|---|---|
| GA_COBB | 2481 | **1324** | 1439 | 69464 | 28722 | **1987** | 143476 | 61743 | **5616** |
| GA_DEKALB | 10638 | 5779 | **1669** | 237839 | 167488 | **10687** | 445521 | 312911 | **9048** |
| GA_DOUGHERTY | **38** | 813 | 2215 | 1398 | **826** | 1365 | **1657** | 1767 | 1877 |
| GA_FULTON | 1167 | 9421 | **782** | 73286 | **10496** | 49138 | 199659 | **16176** | 52948 |
| GA_GWINNETT | 17893 | 11535 | **2273** | 602187 | 474147 | **79634** | 1312633 | 1046657 | **162862** |
| GA_HALL | 38189 | 14280 | **1116** | 405607 | 218666 | **20124** | 723685 | 417116 | **38457** |
| IA_BLACK HAWK | 14356 | 29948 | **884** | 107523 | 149148 | **1617** | 195149 | 242995 | **4701** |
| IA_POLK | 189568 | 6954 | **3664** | 1813782 | 134288 | **39484** | 3189858 | 255139 | **132200** |
| IA_WOODBURY | 1922 | **1758** | 4018 | **5579** | 30735 | 15857 | 27425 | 63053 | **12815** |
| IL_COOK | 5002651 | 4072067 | **210717** | 55305862 | 41072794 | **4411589** | 108187875 | 76314647 | **16615810** |
| IL_DUPAGE | **2309** | 7809 | 5839 | **12249** | 75555 | 92056 | **9748** | 94630 | 451928 |
| IL_KANE | 97118 | 48883 | **10007** | 716906 | 177742 | **28036** | 1529117 | **193514** | 214938 |
| IL_LAKE | 22348 | 20659 | **3121** | 194850 | 150124 | **19432** | 392956 | 268646 | **75696** |
| IL_MCHENRY | 1541 | 971 | **105** | 20823 | 11986 | **2858** | 43753 | 24046 | **2303** |
| IL_WILL | 54643 | **17536** | 28639 | 683068 | 252364 | **195247** | 1302020 | **467553** | 503499 |
| IL_WINNEBAGO | 12008 | 25663 | **637** | 104533 | 63657 | **3476** | 189777 | 89217 | **18648** |
| IN_ALLEN | 3113 | 2207 | **399** | 54806 | 39605 | **11844** | 141443 | 103276 | **27724** |
| IN_CASS | 4281 | 51192 | **161** | 61903 | 244562 | **39374** | 121784 | 373554 | **82848** |
| IN_LAKE | 5620 | 10588 | **94** | 46336 | 88827 | **814** | 97882 | 182628 | **1953** |
| IN_MARION | 114294 | 91430 | **6926** | 1132428 | 892047 | **173817** | 2387723 | 1855274 | **402497** |
| IN_ST. JOSEPH | 1644 | 2559 | **279** | **5614** | 9194 | 7379 | **6187** | 10926 | 9092 |
| KS_FINNEY | 1756 | **413** | 1383 | 6518 | **394** | 87271 | 5355 | **342** | 265359 |
| KS_FORD | 735 | **386** | 1812 | 22771 | 5636 | **947** | 60373 | 15549 | **2929** |
| KS_LEAVENWORTH | 48028 | 11465 | **2948** | 230442 | 32594 | **2929** | 337744 | 42241 | **2240** |
| KS_WYANDOTTE | 5054 | **64** | 222 | 77379 | 8115 | **220** | 156149 | 23969 | **594** |
| KY_JEFFERSON | 10906 | 10230 | **1502** | 85694 | 82463 | **6270** | 220004 | 213616 | **23916** |
| MA_BARNSTABLE | 387 | 752 | **213** | 9777 | 15033 | **934** | 29253 | 40907 | **1074** |
| MA_BRISTOL | 12514 | 51827 | **2235** | **6480** | 102874 | 97789 | **9714** | 181026 | 223140 |
| MA_ESSEX | 151851 | 90157 | **1234** | 1763177 | 1101095 | **4725** | 4300938 | 2854118 | **9959** |
| MA_HAMPDEN | 5641 | 7548 | **4033** | 62516 | 51060 | **48710** | 143037 | 98799 | **60158** |
| MA_MIDDLESEX | 199142 | 55108 | **1296** | 1971598 | 302943 | **19451** | 5544507 | 1129888 | **37912** |
| MA_NORFOLK | 14639 | 22658 | **2028** | 85585 | 85928 | **8694** | 370634 | 326201 | **17925** |
| MA_PLYMOUTH | 24836 | 41936 | **6342** | 156822 | 136391 | **3274** | 460709 | 325898 | **5710** |
| MA_SUFFOLK | 312397 | 332126 | **124583** | 1328344 | 1002098 | **399111** | 3053775 | 2057393 | **609131** |
| MA_WORCESTER | 230221 | 177247 | **4152** | 944588 | 563306 | **151225** | 1838511 | 1041356 | **391841** |
| MD_ANNE ARUNDEL | 2008 | 703 | **429** | 54287 | 25237 | **7824** | 153478 | 75802 | **14848** |
| MD_BALTIMORE | 91553 | 88639 | **18452** | 449595 | 407736 | **67106** | 985647 | 878925 | **126229** |
| MD_FREDERICK | 6510 | 1259 | **1113** | 86343 | 17892 | **12038** | 171214 | 31956 | **18727** |
| MD_HOWARD | 7625 | 7152 | **3344** | 107564 | 98018 | **20722** | 209720 | 188126 | **35380** |
| MD_MONTGOMERY | 179736 | 210302 | **124718** | 1391033 | 1562926 | **413167** | 2938922 | 3121218 | **535881** |
| MD_PRINCE GEORGE'S | 397082 | 579986 | **86136** | 3018645 | 4057905 | **177241** | 6012276 | 7491443 | **142804** |
| MI_GENESEE | 1680 | 1905 | **764** | 7053 | 6005 | **3729** | 13014 | 9712 | **6781** |

| County | | | | | | | | |
|---|---|---|---|---|---|---|---|---|
| MI_KENT | 10762 | 32952 | **621** | 113605 | 278827 | **25423** | 223533 | 506274 | **83847** |
| MI_MACOMB | **991** | 13931 | 6198 | **518** | 51568 | 4545 | **591** | 103669 | 6493 |
| MI_OAKLAND | 12869 | 22483 | **4005** | 28394 | 89500 | **23302** | 47459 | 166421 | **30410** |
| MI_WASHTENAW | **17** | 26 | 316 | **21** | 238 | 253 | **19** | 566 | 605 |
| MI_WAYNE | 83269 | 238758 | **7091** | **144660** | 783108 | 250736 | **202432** | 1294491 | 360187 |
| MN_HENNEPIN | 232924 | 471990 | **8277** | 2828055 | 1748628 | **1232435** | 5721830 | 2512438 | **974097** |
| MN_NOBLES | 4361 | 7410 | **50** | 71850 | 65120 | **15335** | 153882 | 123290 | **31593** |
| MN_RAMSEY | 72683 | 139518 | **472** | 750768 | 592492 | **243283** | 1441383 | 896080 | **223839** |
| MN_STEARNS | 99819 | 8391 | **522** | 714423 | **29040** | 61067 | 1100175 | **38931** | 202698 |
| MO_ST. LOUIS | 25750 | 23919 | **888** | 238588 | 229732 | **25322** | 501347 | 487804 | **38271** |
| NC_DURHAM | 1330 | 4449 | **594** | 41014 | 89773 | **26102** | 109789 | 220458 | **73622** |
| NC_MECKLENBURG | 20347 | 23153 | **18660** | **212885** | 233534 | 323133 | **590732** | 638547 | 864899 |
| NC_WAKE | 1584 | 6970 | **381** | 17082 | 70686 | **14369** | 44968 | 162871 | **38007** |
| ND_CASS | **5048** | 9190 | 5305 | **14637** | 34727 | 50176 | **11731** | 36265 | 51899 |
| NE_DAKOTA | 19069 | 2246 | **528** | 69683 | 77939 | **1012** | 117925 | 152069 | **798** |
| NE_DOUGLAS | 65806 | 122970 | **3088** | 725611 | **473570** | 1669426 | 1829956 | **939132** | 2662947 |
| NE_HALL | 9019 | 10992 | **4240** | 59916 | 64759 | **14255** | 117316 | 121579 | **23566** |
| NH_HILLSBOROUGH | 3958 | 967 | **903** | 54748 | **496** | 26902 | 143957 | **2444** | 26357 |
| NH_ROCKINGHAM | 9667 | 8885 | **3931** | 62103 | 58838 | **1455** | 108403 | 103667 | **1163** |
| NJ_ATLANTIC | 5606 | 7598 | **1060** | 74334 | 68545 | **2620** | 174340 | 144191 | **2462** |
| NJ_BERGEN | 57602 | 43609 | **14097** | 749463 | 432140 | **38632** | 1562395 | 833936 | **47132** |
| NJ_BURLINGTON | 20886 | 7370 | **5470** | 359663 | 199200 | **12902** | 742457 | 443681 | **39671** |
| NJ_CAMDEN | 74153 | 32353 | **16478** | 770110 | 359467 | **4550** | 1594326 | 776531 | **18626** |
| NJ_CUMBERLAND | 28008 | **6203** | 19126 | 260015 | **39695** | 53378 | 536407 | 86277 | **72420** |
| NJ_ESSEX | 314563 | 250949 | **23589** | 1895021 | 1464297 | **41255** | 3590038 | 2806264 | **31434** |
| NJ_GLOUCESTER | 10695 | 10789 | **814** | 95772 | 87840 | **1827** | 183775 | 163840 | **10935** |
| NJ_HUDSON | 570528 | 148403 | **49027** | 2686119 | **235003** | 3990343 | 5034222 | **344182** | 7551543 |
| NJ_MERCER | 121380 | 67124 | **12611** | 1168866 | 753970 | **33796** | 2278538 | 1553999 | **41029** |
| NJ_MIDDLESEX | 184452 | 34017 | **21067** | 2220576 | 586433 | **1778** | 4615178 | 1322828 | **8728** |
| NJ_MONMOUTH | 1204 | 14276 | **420** | 38122 | 109691 | **20080** | 90151 | 202365 | **25572** |
| NJ_MORRIS | 5817 | 7434 | **214** | 87317 | 61142 | **7477** | 229380 | 145189 | **8670** |
| NJ_OCEAN | 38914 | 10048 | **657** | 449643 | 91232 | **500** | 943645 | 183766 | **759** |
| NJ_PASSAIC | 198220 | 98352 | **48633** | 1878404 | 934699 | **116003** | 3866673 | 1965730 | **146634** |
| NJ_SOMERSET | 9485 | **2282** | 3035 | 118735 | 33769 | **475** | 242719 | 70106 | **1943** |
| NJ_UNION | 230920 | 142931 | **11723** | 1201539 | 478223 | **29120** | 2486633 | 932038 | **35648** |
| NM_BERNALILLO | 3775 | 2404 | **441** | 39977 | 24274 | **342** | 84491 | 51512 | **275** |
| NM_MCKINLEY | 12433 | 12249 | **221** | 78035 | 68149 | **45594** | 163865 | 137964 | **57562** |
| NM_SAN JUAN | 2068 | **52** | 375 | 34109 | **547** | 32519 | 94078 | **1259** | 45719 |
| NV_CLARK | **15571** | 27908 | 42104 | 100213 | 179330 | **30981** | 177595 | 317077 | **34806** |
| NV_WASHOE | 9830 | 9690 | **3495** | 31980 | 32303 | **7311** | 57602 | 59071 | **11465** |
| NY_ALBANY | 2118 | 13031 | **981** | 11220 | 84838 | **3900** | 16341 | 141664 | **3328** |
| NY_DUTCHESS | 12570 | 16050 | **5295** | 51028 | 69479 | **21483** | 81019 | 111836 | **21629** |
| NY_ERIE | 28411 | 26691 | **1294** | 561542 | 529422 | **28862** | 1103128 | 1032425 | **27138** |
| NY_MONROE | 29634 | 36475 | **11366** | **66258** | 86075 | 71718 | **70156** | 94745 | 82125 |

| County | | | | | | | | |
|---|---|---|---|---|---|---|---|---|
| NY_NASSAU | 40901 | 222796 | **5613** | 553854 | 1525527 | **259827** | 1378727 | 3054513 | **415975** |
| NY_NEW YORK | **4481900** | 13736222 | 5796699 | 38997234 | 119978471 | **732772** | 86293738 | 252562616 | **841299** |
| NY_ONONDAGA | 7404 | **5925** | 8277 | 90287 | **42704** | 74655 | 165223 | **60091** | 93781 |
| NY_ORANGE | 10580 | 68098 | **612** | 114560 | 561972 | **65925** | 247114 | 1101615 | **92361** |
| NY_PUTNAM | **374** | 2943 | 775 | 11849 | 36610 | **6936** | 24988 | 66187 | **8877** |
| NY_ROCKLAND | 24740 | 75876 | **2819** | 219648 | 425775 | **63111** | 471005 | 779100 | **91923** |
| NY_SUFFOLK | 34498 | 140823 | **5325** | 264714 | 704596 | **92010** | 757294 | 1477103 | **136150** |
| NY_SULLIVAN | 1647 | 3104 | **82** | 7188 | 10222 | **299** | 14698 | 17193 | **1090** |
| NY_ULSTER | 1498 | 6383 | **136** | 8002 | 38172 | **3758** | 15220 | 69847 | **5105** |
| NY_WESTCHESTER | 96710 | 222021 | **3275** | 995981 | 1321971 | **58596** | 2158805 | 2442246 | **73782** |
| OH_CUYAHOGA | 6715 | 10468 | **2089** | 58702 | 111572 | **34729** | 104111 | 215825 | **48039** |
| OH_FRANKLIN | 24100 | 2330 | **1937** | 555182 | 113725 | **3181** | 1319032 | 285580 | **16555** |
| OH_HAMILTON | 633 | 1475 | **168** | 37592 | 54589 | **19322** | 77150 | 108339 | **21984** |
| OH_LUCAS | **16** | 2611 | 41 | 12000 | **4091** | 9978 | 40863 | **3088** | 20046 |
| OH_MAHONING | 3402 | 686 | **230** | 51850 | 22879 | **95** | 100764 | 49053 | **114** |
| OH_MARION | 3999 | 6761 | **1985** | 17235 | 40475 | **1575** | 24586 | 67858 | **1312** |
| OH_PICKAWAY | **567** | 1498 | 1128 | **8296** | 18504 | 15377 | **11791** | 28936 | 22152 |
| OK_OKLAHOMA | 583 | **274** | 367 | 2141 | 1010 | **305** | 2398 | 1012 | **568** |
| PA_ALLEGHENY | 503 | 1316 | **240** | 5469 | 10714 | **590** | 9353 | 17768 | **583** |
| PA_BERKS | 2883 | 1272 | **569** | 7411 | **6162** | 81962 | 5811 | 6285 | 124454 |
| PA_BUCKS | 3222 | **222** | 3121 | 29363 | **1394** | 62365 | 65786 | **2465** | 156482 |
| PA_CHESTER | 365 | **290** | 322 | **360** | 10184 | 472 | **1899** | 37323 | 3464 |
| PA_DELAWARE | 14767 | 5882 | **2096** | 50456 | **5282** | 231007 | 80729 | **5583** | 530827 |
| PA_LACKAWANNA | 4221 | 4284 | **4064** | 13369 | 13112 | **1705** | 18486 | 17710 | **1969** |
| PA_LANCASTER | 4214 | 4833 | **1204** | 70586 | 73934 | **928** | 125565 | 129860 | **1027** |
| PA_LEHIGH | 577 | **434** | 557 | 1394 | **507** | 510 | 2009 | **943** | 1433 |
| PA_LUZERNE | 106 | **57** | 1187 | 2867 | 1046 | **426** | 6460 | 1848 | **1640** |
| PA_MONROE | 30 | 203 | **28** | **590** | 3477 | 1014 | 1392 | 7438 | **1243** |
| PA_MONTGOMERY | 19107 | 36128 | **13765** | 48071 | 127458 | **23246** | 63607 | 193276 | **43896** |
| PA_NORTHAMPTON | 19449 | 21644 | **3604** | 96572 | 105765 | **3751** | 166719 | 180823 | **3296** |
| PA_PHILADELPHIA | 8900 | 73311 | **5543** | 133723 | 2083975 | **118964** | 389486 | 4888068 | **246007** |
| RI_PROVIDENCE | 73961 | **72896** | 79131 | **48303** | 75328 | 1156807 | 156014 | **73946** | 1406586 |
| SC_GREENVILLE | **953** | 1329 | 2150 | **5768** | 7762 | 7824 | **32151** | 37418 | 35045 |
| SC_RICHLAND | **87** | 1589 | 115 | 898 | 7809 | **529** | 2657 | 15175 | **1129** |
| SD_MINNEHAHA | 54066 | 45168 | **1670** | 329942 | 270958 | **51358** | 552250 | 443211 | **86124** |
| TN_DAVIDSON | 38256 | 22294 | **8664** | 409129 | 223940 | **114675** | 859989 | 471462 | **159944** |
| TN_SHELBY | 17083 | 19805 | **5827** | 282417 | 301841 | **157875** | 680238 | 709492 | **328022** |
| TN_TROUSDALE | 49779 | 5354 | **156** | 256031 | 7716 | **18** | 390612 | 22600 | **36** |
| TX_BEXAR | 71067 | 10834 | **9870** | 406311 | **29116** | 75931 | 810178 | **74296** | 129339 |
| TX_COLLIN | 221 | 88 | **33** | 6610 | **355** | 894 | 20069 | **1939** | 2613 |
| TX_DALLAS | 38824 | 86433 | **9751** | 801790 | 252512 | **183019** | 2605412 | 629107 | **306634** |
| TX_DENTON | 859 | 1303 | **94** | 4339 | 6076 | **1971** | 7774 | 10463 | **4085** |
| TX_EL PASO | 13463 | 22830 | **384** | 80869 | 132910 | **3756** | 114309 | 196954 | **3839** |
| TX_FORT BEND | 11556 | 4067 | **901** | 82542 | 30466 | **2177** | 171286 | 71474 | **1082** |
| TX_HARRIS | 91412 | 42218 | **7786** | 382379 | 284220 | **212653** | 772127 | 715891 | **521219** |
| TX_POTTER | 473038 | 367722 | **308955** | 683673 | 286331 | **237608** | 919281 | 280822 | **183812** |
| TX_TARRANT | 27924 | **3022** | 3077 | 495088 | 78464 | **39756** | 852668 | 119766 | **89502** |
| TX_TRAVIS | 910 | **782** | 797 | 44523 | 41865 | **13562** | 132624 | 126234 | **36121** |
| UT_SALT LAKE | 4448 | 2279 | **364** | 73190 | 33692 | **1322** | 245358 | 124108 | **5439** |
| UT_UTAH | 4073 | 2978 | **34** | 76383 | 63891 | **737** | 190014 | 165528 | **4484** |
| VA_ARLINGTON | 4346 | 3918 | **449** | 53808 | 41769 | **5230** | 102038 | 73135 | **5269** |
| VA_CHESTERFIELD | **283** | 955 | 1041 | **1133** | 5933 | 10927 | **8221** | 26980 | 30916 |
| VA_FAIRFAX | 9448 | **4018** | 31774 | **35393** | 119066 | 302720 | **62463** | 251097 | 363120 |
| VA_HENRICO | 1474 | 4450 | **121** | 17522 | 37374 | **16079** | 48833 | 90051 | **43311** |
| VA_LOUDOUN | 2516 | 10393 | **1920** | 167229 | 275123 | **165805** | 319188 | 505620 | **257788** |
| VA_PRINCE WILLIAM | **407** | 2245 | 40121 | **21976** | 50889 | 240130 | **59936** | 109037 | 332086 |
| WA_KING | 7945 | **1880** | 2126 | 43396 | 2345 | **1767** | 138388 | 17732 | **4438** |
| WA_PIERCE | 666 | 761 | **162** | 13399 | 16374 | **1039** | 34482 | 41971 | **3176** |
| WA_SNOHOMISH | 952 | 1483 | **6** | 14203 | 22925 | **518** | 31739 | 51899 | **852** |
| WA_YAKIMA | **17141** | 39890 | 49137 | **128400** | 190995 | 262414 | **369321** | 446671 | 533134 |
| WI_BROWN | 22808 | 5487 | **188** | 273075 | 109151 | **10094** | 500092 | 219092 | **29410** |
| WI_MILWAUKEE | 82335 | 81164 | **27184** | 776944 | 832194 | **537622** | 1373109 | 1511968 | **985075** |
| WI_RACINE | **7067** | 10089 | 10165 | 94742 | 72718 | **1729** | 173955 | 110921 | **10764** |

In order to check the performance difference statistically, we conduct T-test between each two models. The p-values are shown in below tables:

**Table 10.** P-values for state-level prediction (5-days)

| Model | SIR | SEIR | GRU | ColaGNN | CovidGNN | STAN |
|---|---|---|---|---|---|---|
| SIR | 1.00E+00 | 5.93E-01 | 7.24E-05 | 1.40E-06 | 5.19E-06 | 0.00E+00 |
| SEIR | 5.93E-01 | 1.00E+00 | 1.31E-30 | 9.90E-73 | 1.16E-32 | 0.00E+00 |
| GRU | 7.24E-05 | 1.31E-30 | 1.00E+00 | 8.05E-08 | 1.66E-03 | 5.27E-16 |
| ColaGNN | 1.40E-06 | 9.90E-73 | 8.05E-08 | 1.00E+00 | 8.86E-02 | 9.54E-09 |
| CovidGNN | 5.19E-06 | 1.16E-32 | 1.66E-03 | 8.86E-02 | 1.00E+00 | 2.02E-12 |
| STAN | 5.71E-08 | 0.00E+00 | 5.27E-16 | 2.02E-12 | 9.54E-09 | 1.00E+00 |

**Table 11.** P-values for state-level prediction (15-days)

| Model | SIR | SEIR | GRU | ColaGNN | CovidGNN | STAN |
|---|---|---|---|---|---|---|
| SIR | 1.00E+00 | 4.40E-142 | 1.62E-18 | 2.15E-54 | 3.76E-41 | 0.00E+00 |
| SEIR | 4.40E-142 | 1.00E+00 | 2.23E-03 | 3.96E-19 | 1.27E-07 | 0.00E+00 |
| GRU | 1.62E-18 | 2.23E-03 | 1.00E+00 | 2.06E-02 | 8.51E-01 | 3.46E-12 |
| ColaGNN | 2.15E-54 | 3.96E-19 | 2.06E-02 | 1.00E+00 | 3.23E-05 | 4.08E-20 |
| CovidGNN | 3.76E-41 | 1.27E-07 | 8.51E-01 | 3.23E-05 | 1.00E+00 | 4.38E-24 |
| STAN | 0.00E+00 | 0.00E+00 | 3.46E-12 | 4.08E-20 | 4.38E-24 | 1.00E+00 |

**Table 12.** P-values for state-level prediction (20-days)

| Model | SIR | SEIR | GRU | ColaGNN | CovidGNN | STAN |
|---|---|---|---|---|---|---|

| | SIR | SEIR | GRU | ColaGNN | CovidGNN | STAN |
|---|---|---|---|---|---|---|
| SIR | 1.00E+00 | 2.15E-131 | 1.42E-28 | 1.87E-31 | 7.68E-27 | 0.00E+00 |
| SEIR | 2.15E-131 | 1.00E+00 | 5.56E-10 | 6.65E-14 | 4.40E-09 | 4.04E-263 |
| GRU | 1.42E-28 | 5.56E-10 | 1.00E+00 | 7.69E-02 | 9.04E-01 | 2.70E-08 |
| ColaGNN | 1.87E-31 | 6.65E-14 | 7.69E-02 | 1.00E+00 | 6.61E-02 | 1.03E-04 |
| CovidGNN | 7.68E-27 | 4.40E-09 | 9.04E-01 | 6.61E-02 | 1.00E+00 | 5.27E-08 |
| STAN | 0.00E+00 | 4.04E-263 | 2.70E-08 | 1.03E-04 | 5.27E-08 | 1.00E+00 |

Table 13. P-values for county-level prediction (5-days)

| Model | SIR | SEIR | GRU | ColaGNN | CovidGNN | STAN |
|---|---|---|---|---|---|---|
| SIR | 1.00E+00 | 2.30E-13 | 5.52E-03 | 2.67E-10 | 4.43E-04 | 0.00E+00 |
| SEIR | 2.30E-13 | 1.00E+00 | 2.30E-36 | 3.33E-81 | 1.67E-50 | 1.46E-203 |
| GRU | 5.52E-03 | 2.30E-36 | 1.00E+00 | 9.00E-06 | 4.55E-01 | 8.46E-18 |
| ColaGNN | 2.67E-10 | 3.33E-81 | 9.00E-06 | 1.00E+00 | 3.36E-05 | 2.83E-07 |
| CovidGNN | 4.43E-04 | 1.67E-50 | 4.55E-01 | 3.36E-05 | 1.00E+00 | 3.07E-20 |
| STAN | 8.91E-20 | 1.46E-203 | 8.46E-18 | 2.83E-07 | 3.07E-20 | 1.00E+00 |

Table 14. P-values for county-level prediction (15-days)

| Model | SIR | SEIR | GRU | ColaGNN | CovidGNN | STAN |
|---|---|---|---|---|---|---|
| SIR | 1.00E+00 | 2.52E-18 | 1.24E-49 | 2.21E-88 | 4.75E-42 | 0.00E+00 |
| SEIR | 2.52E-18 | 1.00E+00 | 1.25E-78 | 4.86E-116 | 2.29E-71 | 6.21E-253 |
| GRU | 1.24E-49 | 1.25E-78 | 1.00E+00 | 2.22E-12 | 6.08E-01 | 3.57E-84 |
| ColaGNN | 2.21E-88 | 4.86E-116 | 2.22E-12 | 1.00E+00 | 4.00E-09 | 5.23E-60 |
| CovidGNN | 4.75E-42 | 2.29E-71 | 6.08E-01 | 4.00E-09 | 1.00E+00 | 4.32E-69 |
| STAN | 0.00E+00 | 6.21E-253 | 3.57E-84 | 5.23E-60 | 4.32E-69 | 1.00E+00 |

Table 15. P-values for county-level prediction (20-days)

| Model | SIR | SEIR | GRU | ColaGNN | CovidGNN | STAN |
|---|---|---|---|---|---|---|
| SIR | 1.00E+00 | 7.48E-18 | 2.15E-96 | 1.25E-163 | 2.14E-83 | 0.00E+00 |
| SEIR | 7.48E-18 | 1.00E+00 | 3.41E-123 | 1.69E-166 | 5.02E-111 | 9.74E-257 |
| GRU | 2.15E-96 | 3.41E-123 | 1.00E+00 | 6.44E-06 | 6.19E-02 | 5.28E-56 |
| ColaGNN | 1.25E-163 | 1.69E-166 | 6.44E-06 | 1.00E+00 | 1.40E-10 | 9.85E-57 |
| CovidGNN | 2.14E-83 | 5.02E-111 | 6.19E-02 | 1.40E-10 | 1.00E+00 | 2.57E-61 |
| STAN | 0.00E+00 | 9.74E-257 | 5.28E-56 | 9.85E-57 | 2.57E-61 | 1.00E+00 |

The results indicate the performance of STAN have significant difference with baseline models (p-value << 0.05), and STAN performs statistically better than all baselines.